\documentclass{article}
\usepackage{hyperref}
\usepackage{graphicx}
\usepackage[utf8]{inputenc}
\usepackage[T1]{fontenc}
\usepackage{amsmath}
\usepackage{caption}
\usepackage{authblk}
\usepackage{multirow}
\usepackage{amssymb}
\usepackage{geometry}
\usepackage{subfigure}
\usepackage{subcaption}
\usepackage{float}
\usepackage{booktabs} 
\usepackage{colortbl} 
\usepackage{xcolor} 

\title{Anisotropic anomalous diffusion and nonequilibrium in microgravity dusty plasma. Part Two: Spectral Analysis}

\author[1]{Bradley R. Andrew}

\author[1]{Luca Guazzotto}

\author[2]{Lorin S. Matthews}

\author[2]{Truell W. Hyde}

\author[1]{Evdokiya G. Kostadinova}

\affil[1]{Interdisciplinary Plasma Laboratory, Auburn University, Auburn, Alabama, USA}
\affil[1]{Baylor University, Waco , Texas, USA}

\date{}

\begin{document}
\maketitle

\begin{abstract}
Anisotropic anomalous dust diffusion in microgravity dusty plasma is investigated using experimental data from the Plasmakristall-4 (PK-4) facility on board the International Space Station. The PK-4 experiment uses video cameras to track individual dust particles, which allows for the collection of large amounts of statistical information on the dust particle positions and velocities. In Part One of this paper, these statistics were used to quantify anomalous dust diffusion caused by anisotropies in the plasma-mediated dust-dust interactions in PK-4. Here we use scaling relations to convert statistical parameters extracted from data into input parameters for a Hamiltonian spectral model. The kinetic energy term of the Hamiltonian (modeling anomalous diffusion) is informed from the dust displacement distribution functions, while the potential energy term (modeling stochasticity) is informed from fluctuations in the dust positions. The spectrum of energy states for each Hamiltonian is studied to assess probability for extended states (i.e., a continuous portion of the spectrum). The spectral model shows that the combination of nonlocality and stochasticity leads to high probability for transport at certain scales in Hilbert space, which coincide with the characteristic spatial scales of dust particle jumps observed in the experiments. Lastly, we discuss how this spectral approach is generalizable to many complex systems, such as electron transport in 2D materials where statistical models are not feasible.

\end{abstract}

\newpage

\section{Introduction}
\label{sec:Introduction}

Complex (or dusty) plasmas are a collection of electrons, ions, neutral particles, and dust grains (typically micro- to nano-meter in size). Dusty plasmas are ubiquitous in astrophysical and space environments, as well as in laboratory settings, both on Earth and in microgravity environments. Due to the complex interactions between the dust particles and the plasma species, dusty plasmas are observed to exhibit various waves, instabilities, and nonlinear dynamics \cite{shukla_survey_2001, Merlino2012}, which makes them a unique analogue system for the study of complex phenomena such as phase transitions, anomalous diffusion, and metastability. The motion of individual dust particles in these systems are visible at the kinetic level, thus allowing for a reconstruction of the entire phase space for the dust particle ensemble. In addition, particle tracking or velocimetry techniques can be used to obtain statistically significant amount of data from dusty plasma experiments in short periods of time. Another advantage of dusty plasma experiments is that they are reasonably simple to build (table-top) and highly controlled, which makes them ideal for deployment in space. 

\vspace{3mm}

Dusty plasmas have been used to study solid-liquid phase transitions \cite{khrapak_fluid-solid_2012,APS2015,BKostadinova2023,Hariprasad2022}, electroheology \cite{ivlev_electrorheological_2010,Ivlev2011,Pustylnik2020}, strong interparticle coupling and long-range interactions \cite{tsytovich_long-range_1997,smith_dusty_2004,Correia2023}, kinetic theories and diffusion properties \cite{arshad_kinetic_2017,petrov_experimental_2005,liu_particle_2018,Feng2010}, and critical phenomena such as melting and crystallization \cite{Hariprasad2022,Feng2010,Joshi2023}, and turbulence \cite{kostadinova_fractional_2021,Sharma2024,Choudhary2024}. As they exhibit many-body effects, dusty plasmas are useful analogue systems for the study of complex systems such as condensed matter \cite{murillo_strongly_2004,kostadinova_fractional_2021} and smart materials \cite{ivlev_first_2008,ivlev_electrorheological_2010,dietz_phase_2021}. Many other physical aspects and applications of dusty plasma have been summarized in several recent overview papers \cite{Beckers2023,choudhary_perspective_2021,Merlino2021}. With all the characteristics described above, dusty plasmas are ideal for testing new analytical models, especially ones related to nonequilibrium statistical mechanics, anomalous diffusion, and stochasticity \cite{kostadinova_fractional_2021,kostadinova_delocalization_2017,kostadinova_physical_2016,kostadinova_transport_2018,liu_non-gaussian_2008,liu_particle_2018}. In this work, we investigate anomalous diffusion in dusty plasma using a novel spectral approach, where one studies the energy spectrum of an Anderson-type Hamiltonian with a non-local Fractional Laplace operator and a stochastic disorder potential \cite{kostadinova_fractional_2021,kostadinova_delocalization_2017,kostadinova_physical_2016,kostadinova_transport_2018}.

\vspace{3mm}

Here we analyze data from dusty plasma experiments conducted in the Plasmakristall-4 (PK-4) facility on board the International Space Station, where the microgravity environment allows one to neglect gravity and confinement forces needed to compensate for gravity, thus, focusing on plasma-mediated dust-dust interactions. Recent studies using the PK-4 facility \cite{pustylnik_plasmakristall-4_2016} have been focused on complex fluid phenomena, such as dust ionization waves \cite{Naumkin2021,Zhukhovitskii2022,Pustylnik2022}, dust acoustic waves \cite{Goree2020}, ion density waves \cite{Mendoza2024}, shear flows and flow patterns fluctuations \cite{Liu2021}, formation of liquid crystalline structures \cite{pair_corr_Gehr2025, Kostadinova2023}, and anomalous heat dissipation  \cite{McCabe2025}. Additionally, analysis of PK-4 data has inspired a breadth of numerical studies including particle-in-cell simulation of PK-4 predicting the formation of ionization waves \cite{Hartmann2020} and molecular dynamics simulations of dust and ions investigating how such ionization waves can cause anisotropies in the ion wakefields around the dust grains \cite{Mendoza2025, Vermillion2023, pair_corr_Gehr2025}. The experimental data sets used in the present study were also used in \cite{pair_corr_Gehr2025,McCabe2025}, which allows for meaningful comparison of results.

\vspace{3mm}

The PK-4 experiment uses video cameras to track individual dust particles, yielding large amounts of statistical information on the dust particles positions and velocities. Previous studies of dusty plasma with the PK-4 experiment have shown velocity distribution functions (VDFs) that were non-Maxwellian with high-energy tails, suggesting anomalous dust diffusion. Anomalous diffusion is a microscopic process which leads to a mean squared displacement (MSD) that grows non-linearly with time $MSD\propto \tau^\alpha$, where $\tau$ is time delay and $\alpha$ is the exponent that quantifies the nonlinearity. If the MSD growth is faster than linear with time, $\alpha>1$, the particles are superdiffusive, while a growth rate slower than linear, $\alpha<1$, indicates subdiffusion. Anomalous diffusion and corresponding non-linear MSDs have been observed in biological molecular transport \cite{tarantino_tnf_2014}, L\'{e}vy flight movements of organisms \cite{reynolds_liberating_2015}, condensed matter physics \cite{benhamou_lecture_2018}, and dusty plasmas \cite{feng_identifying_2010,liu_superdiffusion_2008,liu_non-gaussian_2008}. A review of classical and anomalous diffusion across many fields can be found in \cite{oliveira_anomalous_2019}. Mathematically, anomalous diffusion can be modeled using fractional derivatives and, in particular, the Fractional Laplace operator $(-\Delta)^s$, where $s$ is inversely proportional to the mean squared displacement exponent $\alpha = 1/s$ in the asymptotic limit of $\tau\rightarrow\infty$ In this study, we use a series representation of the fractional Laplacian in the one-dimensional case and study its energy spectrum in Hilbert space.

\vspace{3mm}

The spectral approach presented here is based on the extended states conjecture developed by Liaw \cite{liaw_approach_2013} and the series representation of the fractional Laplacian developed by Padget et al. \cite{fruge_jones_series_2021},\cite{padgett_anomalous_2020}. The physical interpretation of the spectral approach can be found in \cite{kostadinova_physical_2016,kostadinova_delocalization_2017} and its applications to dusty plasma as an analog to condensed matter and materials have been discussed in \cite{kostadinova_transport_2018,kostadinova_fractional_2021}. The core idea of the mathematical approach presented here, the Fractional Laplacian Spectral Method (FLSM), is that the relationship between a given dust-dust interaction potential and the resulting structure or dynamics of the dusty particle system can be inferred from the spectral properties of the corresponding Hamiltonian operator. For a given dusty plasma system, the observed microscopic dynamics (at the individual dust particle scales) can be used to define a Hamiltonian that contains the information of the nonlocal interactions and stochastic fluctuations driving the dust diffusion. Then, the time-evolved dynamics of the dust system can be inferred from the spectrum of the Hamiltonian, i.e., the collection of possible energy states. 

\vspace{3mm}

During a structural transition (here the transition is from a homogeneous to a filamentary state), changes in the structure function and the correlation lengths within the system also yield changes in the corresponding energy spectrum and related diffusion properties. This relationship has previously been used in the study of the metal-to-insulator transition in the Anderson localization problem \cite{anderson_absence_1958,edwards_numerical_1972} and it forms the basis of the scaling theory of Anderson localization \cite{abrahams_scaling_1979}. Here, we will use a similar principle to quantify changes in the energy spectra for different structural states observed in the PK-4 dusty plasma experiments. The key advantage of studying infinite-dimensional operators in Hilbert space is the ability to study the available energy states for a system with given global characteristics without the need to know the details of the local dynamics and without the need to impose boundary conditions \cite{kostadinova_delocalization_2017,kostadinova_physical_2016}. The main requirement for the validity of the proposed spectral method is to ensure that the experiment or the simulation yields enough data so that a meaningful statistical analysis can be used to inform the correct form of the corresponding Hamiltonian. This statistical analysis was performed in Part One of this study \cite{Andrew2025}.

\vspace{3mm}

In Part One of this study, we conducted a statistical analysis of particle tracking data from 9 datasets from PK-4 experiments. Each dataset represented a different combination of pressure-current conditions, which resulted in a different structural state of the dusty plasma cloud. We used fits to dust mean squared displacements (MSDs) and position histograms to obtain parameters that can be scaled as inputs for the Fractional Laplacian in FLSM. Table 1 shows the 9 pressure current datasets, along with the main fitted parameters - the MSD exponents $\alpha$ and the nonextensive parameter $q$, which quantify tailedness of the displacement distribution functions (see Part One of this paper \cite{Andrew2025} for more details). From this information, we can obtain the fraction $s$ on the Laplacian, which will be discussed in Sec.\ref{sec:Theory}. 

\vspace{3mm}

The remainder of this paper is organized as follows. A brief overview of the experimental setup used for the PK-4 experiment is provided in (\autoref{sec:Experiment}). The theoretical background of modeling anomalous diffusion via nonextensive statistics, fractional diffusion equations, the L\'{e}vy distribution, and the Fractional Laplacian Spectral Method (FLSM) is provided in (\autoref{sec:Theory}). The calculation of dimensionless disorder for each structural state is discussed in \autoref{sec:Disorder}, while \autoref{sec:scaling} provides a discussion on the scaling relations used in the study. A summary of results is presented in (\autoref{sec:Analysis}), followed by a discussion in (\autoref{sec:Discussion}). Conclusions and future work are outlined in (\autoref{sec:Conclusion}).

\section{Experimental Setup}
\label{sec:Experiment}

\begin{figure}[H]
    \centering
    \includegraphics[width=80mm]{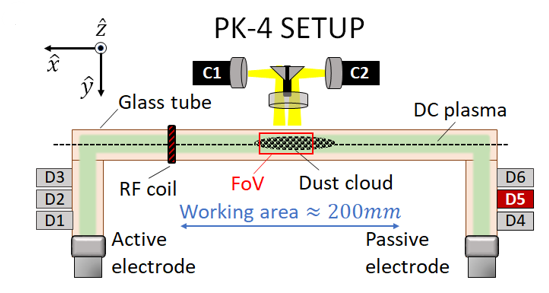}
    \caption{PK-4 Experimental Setup. Dust from Dispenser 5 is trapped in the camera field of view in a DC neon discharge using polarity switching of the DC electric field.}    \label{fig:PK4}
\end{figure}

Here we briefly discuss the PK-4 experimental apparatus \cite{pustylnik_plasmakristall-4_2016}, and the specifics of the Campaign 7 (C7) data that were used for the analysis. The core of the PK-4 facility is comprised of an integrated baseplate housing the diverse components, including a glass plasma chamber with electrodes and microparticle dispensers (injecting Melamine-Formaldehyde dust spheres), vacuum and gas supply systems, plasma generation and diagnostic tools, microparticle manipulation devices, a microparticle observation system with cameras, and an illumination laser, see Figure \ref{fig:PK4}. The main vacuum vessel is a cylindrical glass chamber, where plasma can be created using a direct current (DC) discharge power supply or using radio frequency rf coils. The present experiments were conducted in pure DC neon discharge. Polarity switching of the DC current at different frequencies and variable duty cycles can be used to transport the dust particles (using an asymmetric duty cycle) and capture them in the cameras FoV (using a symmetric duty cycle). Several microparticle manipulation options and on-board plasma diagnostics, such as a plasma glow observation system and a mini spectrometer, are also available. Here we label the axial direction in  PK-4 (x direction in Figure \ref{fig:PK4}) as $\|$ since it is parallel to the direction of the dc electric field. The radial, or cross-field, direction in the camera's field of view (z direction in Figure \ref{fig:PK4}) is labeled $\perp$.

\vspace{3mm}

The DC discharge plasma is generated by two electrodes within the $\pi$-shaped glass chamber. A custom-made bipolar high-voltage (HV) power supply serves as a current source, providing a stabilized output current of up to 3.1 mA at a maximal overall voltage of 2.7 kV. The current is regulated on the active side, with return current measurement on the passive side. Steady-state deviations from the set current value remain below 5$\%$. When a symmetric duty cycle is used with a fast polarity switching of the current (here 500 Hz), the dust microparticles are unable to respond as the dust response frequency is close to 10 Hz. This results in overall stationary negatively charged microparticles suspended within a 'sloshing' stream of ions that forms anisotropic ion wakefields surrounding the dust grains. The Particle Observation (PO) system facilitates microparticle imaging, employing a 532 nm diode laser and two PO cameras with CCD chips of 1600 × 1200 pixels. The cameras are movable and can cover the entire volume of the working area. The camera frame rate used in the present study is 70.1 fps. 

\vspace{3mm}

The PK-4 Campaign 7 experiments discussed here were conducted on July 26, 2019. Our analysis uses the video data from nine sets of pressure-current conditions. The polarity switching frequency was 500 Hz with a duty cycle of $50\%$. Microparticles of size of $3.38 \mu m$ diameter were used. In each case, the dust cloud was allowed to settle for ~50 s, after which a scan of the laser sheet was performed through the dust cloud (along the y-axis in Figure \ref{fig:PK4}). These y-scans allow for obtaining information on the 3D structure of the cloud. The statistical analysis presented here uses particle tracking data collected over a period in which the dust cloud had settled. The cameras and laser sheet are focused on the mid-plane of the cloud. We use the dust particle positions and velocities to study the probability distribution functions, anomalous diffusion, temperature, and nonequilibrium properties.

\section{Theory of Anomalous Diffusion }
\label{sec:Theory}
\subsection{Statistical Approach}

In the absence of long-range interactions or correlations, the particle diffusion can be described by a Brownian motion. The model differential equation is the well-known diffusion equation

\begin{equation}
    \frac{\partial p(x,t)}{\partial t}= \Delta ( D p(x,t)),
    \label{eq:normaldiff}
\end{equation}
 where $\Delta=\frac{\partial^2}{\partial x^2}$ is the Laplacian operator, $D$ is the diffusion constant, and $p(x,t)$ is the distribution function. Equation \ref{eq:normaldiff} is also the linear Fokker-Planck equation with no drift. The one-dimensional solution to the diffusion equation has a Gaussian distribution functional form given by

\begin{equation}
    p(x,t)=\frac{1}{\sqrt{2\pi Dt}} e^{-\frac{(x-x_0)^2}{2Dt}}.
    \label{eq:diffusioneq}
\end{equation}

The mean squared displacement (MSD) is the second moment of this equation: $\langle(x-x_0)^2\rangle\equiv \int (x-x_0)^2 p(x,t)dx=Dt$. As can be seen, the classical diffusion equation yields MSD that grows linearly with time. Given discrete particle positions, the MSD expression for $N$ particles can be generalized in two or three dimensions to

\begin{equation}
    \langle |r(t)-r_0|^2\rangle = \frac{1}{N}\sum_{i=1}^N |r^{(i)}(t)-r^{(i)}(t=0)|^2 = 2dDt^\alpha   \hspace{4mm}  \alpha \geq 0, 
    \label{eq:genmsd}
\end{equation}

Here $d$ is the dimension and $\alpha$ is an exponent that quantifies deviations from a linear dependence in time (i.e., the presence of anomalous diffusion). When $\alpha=1$ (linear MSD), the diffusion is classical, and $\alpha\neq 1$ implies anomalous diffusion. While anomalous diffusion is ubiquitous in nature  \cite{tarantino_tnf_2014, reynolds_liberating_2015, benhamou_lecture_2018,feng_identifying_2010,liu_superdiffusion_2008,liu_non-gaussian_2008,oliveira_anomalous_2019}, its mathematical description is challenging and requires different mathematical models than Eq. \ref{eq:normaldiff}.

\vspace{3mm}

One way to study anomalous diffusion is with a nonlinear diffusion equation, such as the nonlinear Fokker-Planck equation, which has several exact solutions including Porous Medium solutions (see page 249 of \cite{frank_nonlinear_2005}) and Tsallis $q$-Gaussian distributions \cite{tsallis_introduction_2009} (also known as Kappa distributions or Student's $t$ distributions). Anomalous diffusion may also be studied by use of fractional spatial derivatives \cite{kostadinova_delocalization_2017,kostadinova_transport_2018,padgett_anomalous_2020,kostadinova_fractional_2021} which lead to solutions such as L\'{e}vy distributions. We note that it is not well understood if universal scaling relations exist that link the different proposed analytical formulations for anomalous diffusion. While there are some direct relations between the nonlinear Fokker-Planck equations and the $q$-Gaussian solutions, their relation to fractional differential equations is not straightforward even though they can both describe similar regimes of anomalous diffusion. The nonlinear Fokker-Planck equation with no drift

\begin{equation}
\frac{\partial p(x,t)}{\partial t}=D\frac{\partial^2 [p(x,t)]^\nu}{\partial x^2}
\label{eq:nonlindiff}
\end{equation}
has the explicit solution \cite{frank_nonlinear_2005}

\begin{equation}
p(x,t)=[2D(q+q^2)(t-t_0)A_q]^{\frac{-1}{1+q}}*\left[1+\frac{(1-q) (x-x_0 )^2}{4[(2D(q+q^2)(t-t_0)A_q^{1-q})^{\frac{2}{1+q}}]}\right]^{\frac{-1}{1-q}}.
\label{eq:Dq}
\end{equation}
with $\nu \equiv 2-q$ and may be further simplified as shown in \cite{tsallis_introduction_2009}. $A_q$ is a normalization factor given by

\begin{equation}
A_q= 
\begin{cases}\sqrt{\frac{\pi}{(q-1)}}\frac{\Gamma(\frac{q}{q-1})}{\Gamma(\frac{3-q}{2q-2})} & \text{if } 1<q<3 \\                                  \sqrt{\pi}  & \text{if } q=1 \\
 \sqrt{\frac{\pi}{(1-q)}} \frac{\Gamma(\frac{1+q}{2-2q})}{\Gamma(\frac{1}{1-q})} & \text{if } q<1.	
\end{cases}
\label{eq:qnorm}
\end{equation}

The parameter $q$ can also be related to the distribution kurtosis $\kappa$, through the relation: $\kappa = \frac{15-9q}{7-5q}$, for $q<7/5$ \cite{q2kurtSalah} where $k=\frac{\langle x^4\rangle}{\langle x^2\rangle^2}$. Kurtosis has been used in other studies to quantify the deviation from a Gaussian distribution and the related anomalous diffusion (e.g., as discussed in \cite{Ghannad2021}). The primary interest of the present study is the range $-1<q<3$. For $1<q<3$ the distributions exhibit leptokurtic or ‘fat-tail’ behavior. This regime is also where anomalous diffusion is superdiffusive ($\alpha >1$). For $-1<q<1$ the distributions exhibit platykurtic  or 'flat peaks' and weak tails and are subdiffusive cases where $\alpha<1$. The scaling between MSD and $q$-Gaussians can be seen by taking the second moment of Eq. \ref{eq:Dq}, which yields a non-linear relation between the MSD and time

\begin{equation}
\langle (x-x_0)^2 \rangle \propto D_qt^{\frac{2}{3-q}} ;   \hspace{6mm} q<\frac{5}{3},	
\label{eq:alpha2q}
\end{equation} 
where $D_q$ is a new diffusion coefficient. 

\vspace{3mm}

Nonextensive statistics were proposed by Tsallis \cite{tsallis_possible_1988} and are described fully in the textbook on the subject \cite{tsallis_introduction_2009}. The probability distribution function in this formulation is called a $q$-Gaussian, where the parameter $q$ quantifies the nonextensivity of the system. The nonextensivity feature of the formulation makes it independent of initial conditions and suitable for modeling many-body complex systems with long-range interactions, such as plasma. Systems with this nonextensive behavior no longer have additive entropies, $S_q(A+B)\neq S_q(A) + S_q(B)$, and the different microstates are not in equilibrium. Therefore, quantifying the nonextensivity is needed to make rigorous statistical claims and understand equilibrium properties of systems with long-range interactions and correlations. Thus, $q$ can be understood as a parameter quantifying the strength of correlations or non-local interactions causing the system to move away from equilibrium. The nonextensive parameter $q$ can be found by extremizing the Tsallis entropy $S_q$ with Lagrange constraints, such as normalization and moments of the distribution. The variances of these distributions are finite for $q<5/3$ but diverge for $5/3\leq q<3$. This means that the distribution approaches a Gaussian for $q\rightarrow1$, while, for $5/3\leq q<3$, it is a L\'{e}vy type distribution. For the position displacement distributions $q_p>5/3$, the diffusion is a L\'{e}vy process which yields large 'jumps' or L\'{e}vy flights. According to Eq. \ref{eq:alpha2q}, there is no relationship between a L\'{e}vy distribution and a $q$-Gaussian if $q<\frac{4+d}{2+d}$, where $d$ is the dimensionality of the system. Thus, for $d=1$,  $q<\frac{5}{3}$ (implying the mathematical variance is finite), there are no L\'{e}vy processes or L\'{e}vy flights.

\subsection{Connecting the Statistical Approach to the Spectral Approach}

A L\'{e}vy distribution can be found from the following fractional partial differential equation \cite[Eq.~(34), p.~15; Eq.~(58), p.~26]{metzler2000random, podlubny1999fractional}:

\begin{equation}
\frac{\partial p(x,t)}{\partial t}=D \frac{\partial^\gamma p(x,t)}{\partial|x|^\gamma}, \quad 0<\gamma<2.
\label{eq:fractionalpdf}
\end{equation}
The solution in one dimension is known as a L\'{e}vy distribution \( L_{\gamma}(x) \) \cite[Eq.~(61), p.~33]{metzler2000random}:

\begin{equation}
    p(x,t) = (Dt)^{1/\gamma} L_{\gamma} \left( \frac{x}{(Dt)^{1/\gamma}} \right).
\end{equation}

Typically, \( \alpha \) is used and these are called \( \alpha \)-stable L\'{e}vy processes. However, since we use \( \alpha \) for the power-law exponent in the mean squared displacement (MSD), we adopt the symbol \( \gamma \). The index \( \gamma \) relates to the order \( s \) of the fractional Laplacian operator by \( \gamma = 2s \) \cite[Chapter~5, pp.~222--224]{podlubny1999fractional}, and to the nonextensive parameter \( q \) through \( \gamma = \frac{3 - q}{q - 1} \). The MSD of a L\'{e}vy process is given by the relation $\langle x^2 \rangle \propto t^{2/\gamma}$, in the asymptotic limit of large $t$. Since $\gamma$ is defined in the range $0<\gamma<2$, this process is classified as superdiffusive. A L\'{e}vy distribution does not obey the classical central limit theorem, as it has infinite variance. Instead, it arises from the generalized central limit theorem, which applies to sums of independent identically distributed (i.i.d.) random variables with heavy tails. If the individual variables are instead constrained to have finite variance, their sum converges to a Gaussian distribution by the classical central limit theorem. Thus, one of the unique properties of L\'{e}vy distributions is that the random variables do not need to have a finite variance. Infinite variance would seem to imply an infinite second moment, which corresponds to infinite diffusion or temperature depending on the integrated space. This,  of course, is not physical. The solution lies in taking a closer look at the step size of the random walk that yields the L\'{e}vy distribution. During a L\'{e}vy process, a "L\'{e}vy flight" is an occasional occurrence of a very large step size or "flight", while most step sizes are small. In the mathematical definition, there is no restriction to the size of this jump, i.e. no boundary conditions. In a standard random walk that leads to a Gaussian distribution, the step size is proportional to the mean free path $\lambda_{mfp}$. However, we propose that when a system exhibits correlations and long-range interactions, there will exist other subsets of particle displacements, where the size of the subset decreases with increasing $q$ and the length of the step-size increases with increasing $q$. We will call $\lambda_q > \lambda_{mfp}$ "large jumps", and as $\lambda_q \ggg \lambda_{mfp}$ this tends toward L\'{e}vy flights. While L\'{e}vy flights are scale invariant and have infinite variance mathematically, physically they must be smaller than some upper bound  $\lambda_q < L$, where $L$ is based on the system size. Therefore, a L\'{e}vy process is a subset of superdiffusion, which is a subset of anomalous diffusion processes, one in which some step-sizes are much greater than the rest. It is perhaps possible that these cascading scales of step-sizes have a fractal nature, which corresponds with much of the research surrounding Tsallis statistics. 

\vspace{3mm}

The relationship between the fractional differential equations and fractional linear operators can be seen when Eq.\ref{eq:fractionalpdf} is rewritten in terms of the 1D Fractional Laplacian

\begin{equation}
    \frac{\partial p(x,t)}{\partial t}=D (-\Delta)^s p(x,t)
    \label{eq:fraclpdf}
\end{equation}
for values of $0<s<1$, this describes a superdiffusive process. For $s>1$, the fractional Laplacian formulation also models a subdiffusive process. However,  the upper cutoff on the validity of this is not certain, especially since $s=2$ yields the Bi-harmonic operator.

\vspace{3mm}

\begin{figure}[H]
    \centering
    \includegraphics[width=100mm]{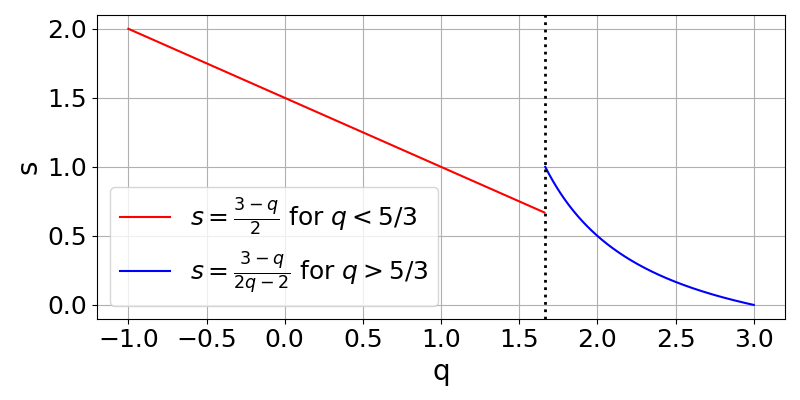}
    \caption{Scaling relations between the nonextensive parameter $q$ and the fraction $s$ on the Laplacian operator for two distinct regimes. The critical value separating the regimes is $q=5/3$.}    
    \label{fig:svq}
\end{figure}

The relation between Tsallis's $q$ and the fraction $s$ can be found in the context of L\'{e}vy distributions and the shared relationship with the MSD \cite{tsallis_introduction_2009}. This yields

\begin{equation}
    \frac{1}{\alpha} = s=\frac{3-q}{2}
    \label{eq:s2qlongtime}
\end{equation}
in the long time delay limit $\tau\rightarrow\infty $ with $q<5/3$, and

\begin{equation}
    s=\frac{3-q}{2q-2} \hspace{2mm} \text{for} \hspace{2mm} q\geq 5/3.
    \label{eq:s2qlarge}
\end{equation}
These relations are plotted in Figure \ref{fig:svq}.

\begin{figure}[H]
    \centering
    \includegraphics[width=140mm]{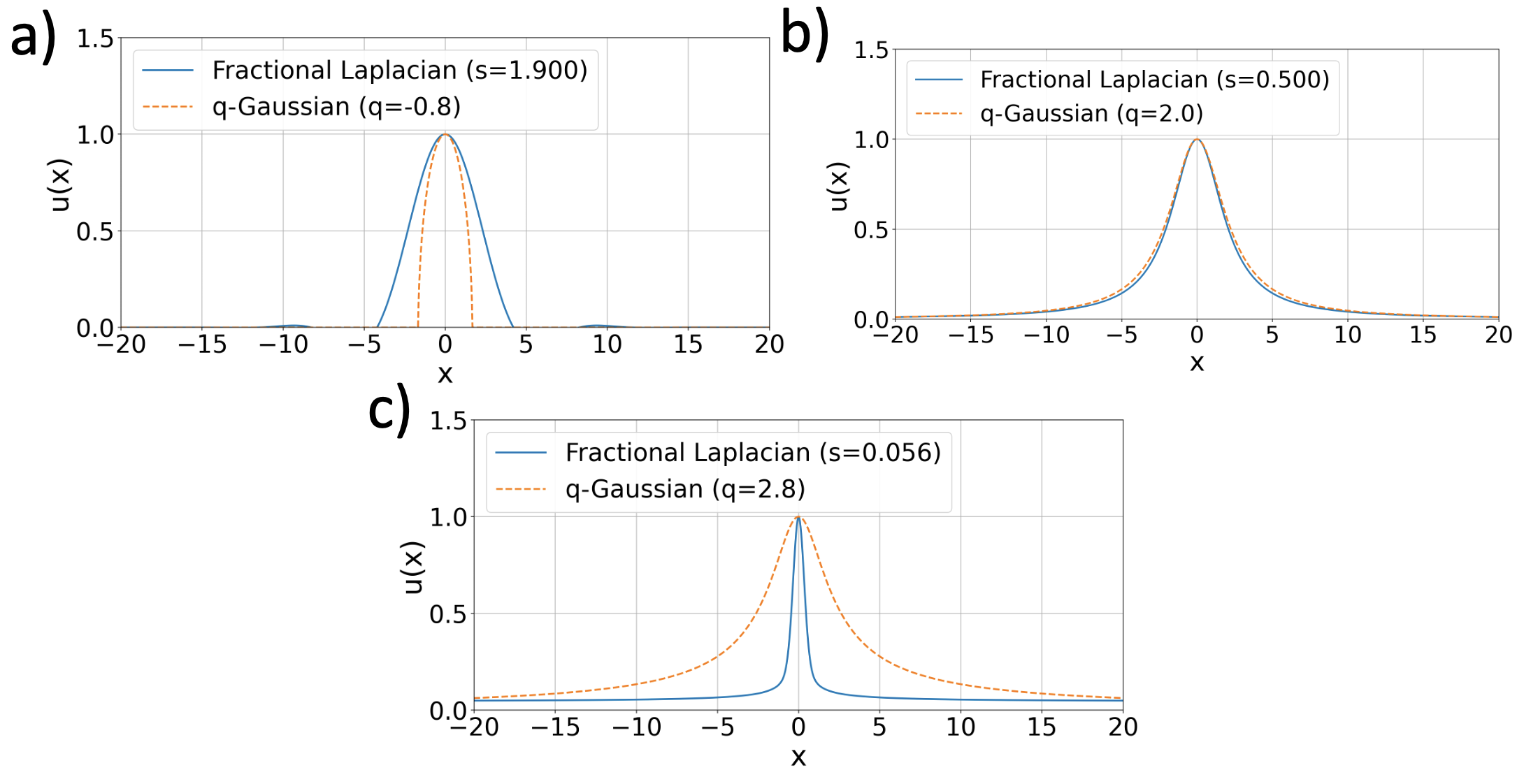}
    \caption{Fractional Laplacian diffusion solution (solid) and $q$-Gaussian solution (dashed) for a) $q=-0.8$ (subdiffusion case $s=1.9$), b) $q=2.0$ (L\'{e}vy flight case $s=0.5$), and c) $q=2.8$ (L\'{e}vy flight case $s=0.056$).}
    \label{fig:qgauss_comparison}
\end{figure}

The discontinuity at $q=5/3$ means that there is not a bijective mapping between $s$ and $q$. The difference between the two scaling relations at $q=5/3$ raises questions on which model is more valid in the range $2/3<s<1$. If one measures the MSD of a system and uses $\alpha = 1/s$ to find $s$, then either $s=\frac{3-q}{2}$ or $s=\frac{3-q}{2q-2}$ can be used to solve for $q$ in the range $2/3<s<1$. However, solving for $q$ from $\alpha$ yields only one $s$, from either Eq. \ref{eq:s2qlongtime} or \ref{eq:s2qlongtime}. A Fast Fourier Transform (FFT) numerical solution of Equation \ref{eq:fraclpdf},  ($p_s$), is compared to the solution of the $q$-Gaussian diffusion Eq. \ref{eq:Dq} ($f_q$) with a sufficiently long time delay after the initial condition for three representative cases in Figure \ref{fig:qgauss_comparison}.

\begin{figure}[H]
    \centering
    \includegraphics[width=90mm]{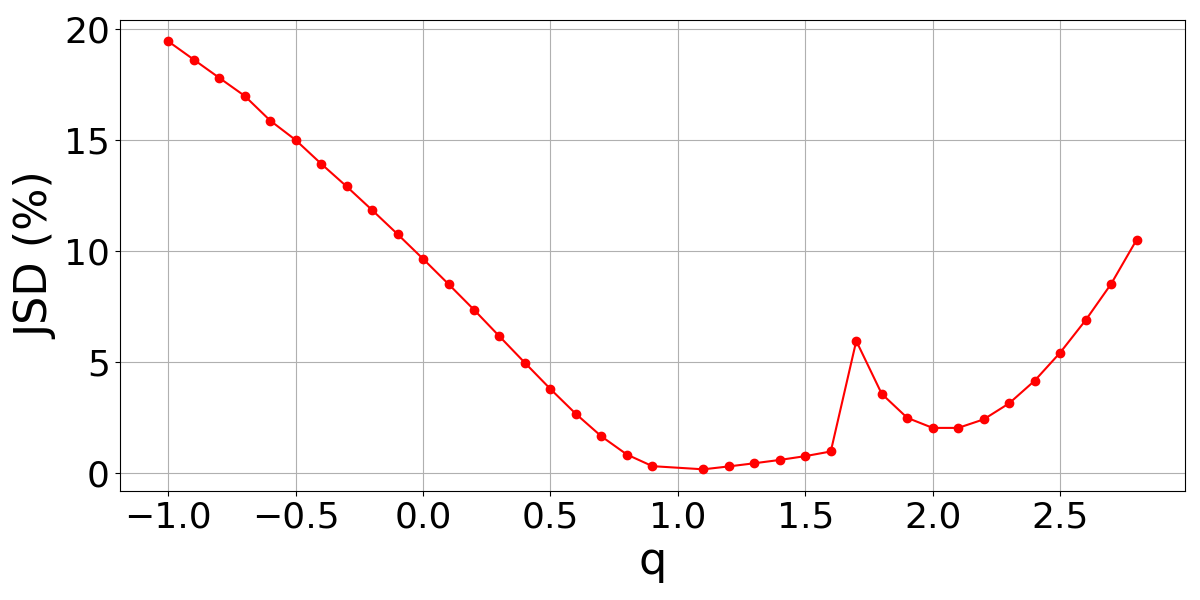}
    \caption{Jensen--Shannon Divergence  across a range of $q$ values. JSD drops rapidly near $q=1$}
    \label{fig:jsd}
\end{figure}

The three figures are representative of two extremes: $q=-0.8, s=1.9$ and $q=2.8, s=0.056$ and the intermediate case of $q=2, s=0.5$, where the case $s=0.5$ has been well studied mathematically. As can be seen in Figure \ref{fig:qgauss_comparison}, the two solutions converge well for the intermediate case and substantially differ for the extreme cases. To assess the similarity between the numerical solution $p_s(x)$ and the $q$-Gaussian profile $f_q(x)$, we computed two key metrics: the Jensen-Shannon divergence (JSD) and the pointwise residual across $x$ and $q$. The JSD is a symmetric and bounded variant of the Kullback--Leibler (KL) divergence, which is used for comparing two probability distributions.

\begin{equation}
   D_{\mathrm{JSD}}(p_s \| f_q) = \frac{1}{2} D_{\mathrm{KL}}(p_s \| m) + \frac{1}{2} D_{\mathrm{KL}}(f_q \| m), \quad \text{where } m = \frac{1}{2}(u + f_q). 
\end{equation}
The KL divergence itself is defined as

\begin{equation}
    D_{\mathrm{KL}}(P \| Q) = \sum_i P(x_i) \log \left(\frac{P(x_i)}{Q(x_i)}\right),
\end{equation}
and measures the information lost when $Q$ is used to approximate $P$. It is asymmetric and can diverge if $Q(x_i) = 0$ while $P(x_i) > 0$. The JSD overcomes these limitations by symmetrization and smoothing the distributions. In our case, the JSD quantifies how much the numerical PDF $p_s(x)$ diverges from the theoretical $q$-Gaussian $f_q(x)$. JSD is typically bounded between $0$ and $\ln{(2)}$, but we rescale to a percentage between $0\%$ and $100\%$. High values of JSD for $q < 1$ indicate poor agreement between the two (e.g., due to cutoff behavior or long tails), while its sharp decrease near $q=1$ reflects strong similarity. As shown in Figure \ref{fig:jsd}, the JSD under 10\%for $0<q<2.7$, indicating close agreement in these anomalous diffusion regimes.

\begin{figure}[H]
    \centering
    \includegraphics[width=120mm]{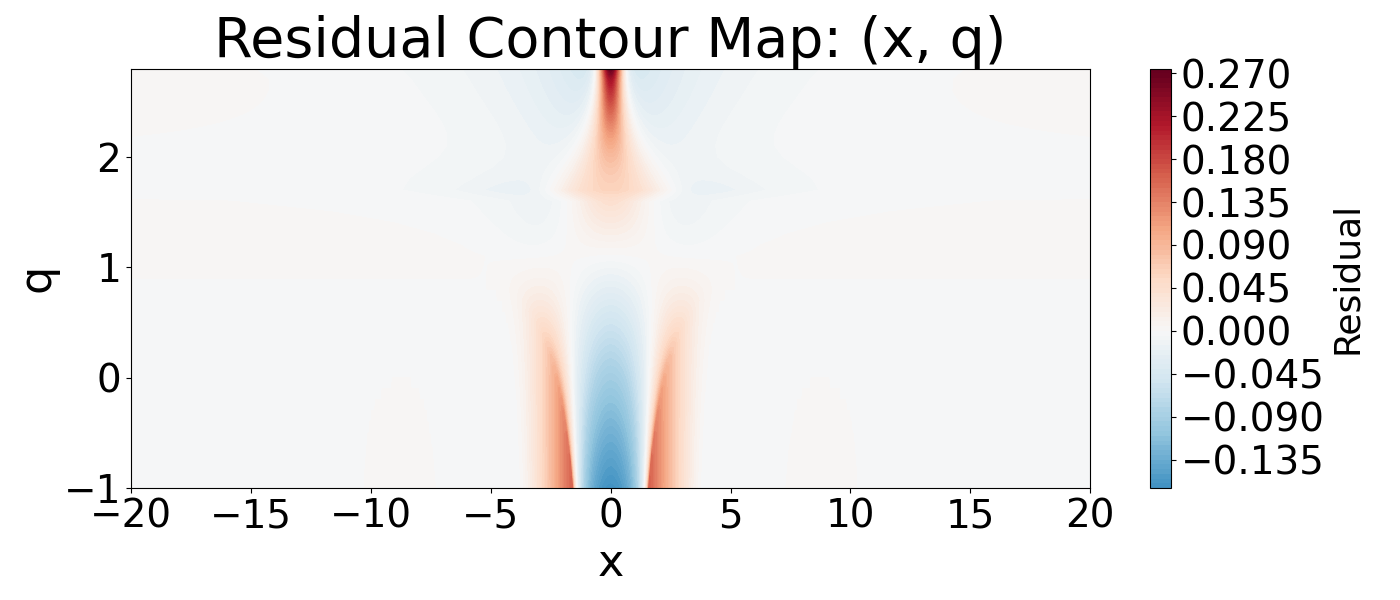}
    \caption{Residual contour plot showing the pointwise difference $\Delta(x, q) = p_s(x) - f_q(x)$ across the spatial domain $x$ and parameter $q$. Blue regions indicate excess in the fractional Laplacian solution, red regions indicate excess in the $q$-Gaussian.}
    \label{fig:res}
\end{figure}

To understand the spatial nature of the error, we computed a residual:
\[
\Delta(x, q) = u(x) - f_q(x),
\]
and plotted it across $x$ and $q$. The contour map, shown in Figure \ref{fig:res}, highlights for which $q$ values the deviations are most significant.

The $q$-Gaussian solution corresponds to the nonlinear diffusion Equation \ref{eq:nonlindiff}, while the $p_s(x)$ solution corresponds to the fractional diffusion Equation \ref{eq:fraclpdf}. There seem to be parameter spaces where the two agree both in the bulk and tails of the distribution. These are the white regions in Figure \ref{fig:res}. This brings the argument that one may model systems with nonlinear dynamics with fractional operators if the appropriate scaling is known. For the above example, it seems that the range $1/2<q<2.1$ or $0.41<s<1.25$ exhibits small enough differences between the solutions of the two diffusion equations, suggesting that scaling between $q$ and $s$ can be used to connect nonextensive statistics to the spectral model. Next, we discuss the spectral approach which makes use of the Fractional Laplacian.

\subsection{Spectral Approach}

In addition to nonlocal interactions and correlations, anomalous diffusion can be affected by stochastic processes, such as random fluctuations of the potential energy across the spatial domains. This is especially true for strongly coupled systems, like dusty plasmas, where the electrostatic interaction potential is sensitive to both fluctuations in the dust charging and the relative position of each dust grain in space. Here we combine the effects of nonlocality and stochasticity by studying a discrete random fractional Schrodinger operator. This Anderson-type Hamiltonian has a fractional Laplacian kinetic energy term and a stochastic disorder potential energy term.  The energy spectrum of this Hamiltonian has been recently used by Kostadinova et al. to determine the onset of turbulence in dusty plasma monolayers \cite{kostadinova_fractional_2021}. As discussed in the previous sections, here we use the series representation of the Fractional Laplacian, first introduced by Padgett et al. \cite{padgett_anomalous_2020} and study the energy spectrum of the Hamiltonian using the extended states conjecture derived by Liaw \cite{liaw_approach_2013}. 

\vspace{3mm}

The discrete random fractional Schrodinger (DRFS) operator has the following form

\begin{equation}
H_{s,\epsilon}\equiv (-\Delta)^s + \sum \epsilon_i \langle \cdot,\delta_i \rangle \delta_i
\label{eq:H}
\end{equation}
where the fractional Laplacian operator $ (-\Delta)^s, s\in(0,2)$  is defined as

\begin{equation}
    (-\Delta)^s u_{n} \equiv \sum_{m\in\mathbb{Z}}^{m\neq n} K_s(n-m)(u_n - u_m) 
\end{equation}

\begin{equation}
    K_s(m)\equiv \begin{cases} 
    \frac{4^s \Gamma(1/2+s)}{\sqrt{\pi}|\Gamma(-s)|}\frac{\Gamma(|m|-s)}{\Gamma(|m|+1+s)} & \text{if } m\in\mathbb{Z}/\{0\} \\
    0 & \text{if } m=0
    \end{cases}
    \label{kernel}.
\end{equation}
The kernel  $K_s(m)$ in the above expression serves as a weight that quantifies the range and strength of the nonlocal interaction (i.e., interaction beyond the nearest neighbors). The potential energy term in Eq. \ref{eq:H} consists of independent variables $\epsilon_i$, identically distributed according to a uniform (flat) distribution on the interval $[-c/2,c/2]$, where $c>0$ is a dimensionless disorder parameter. In equation \ref{eq:H}, $\delta_i$ is the $i^{th}$ standard basis vector of the 1D integer space $\mathbb{Z}$ and $\langle \cdot, \cdot \rangle$ is the $l^2(\mathbb{Z})$ inner product in the space.

\vspace{3mm}

\begin{figure}[H]
    \centering
    \includegraphics[width=140mm]{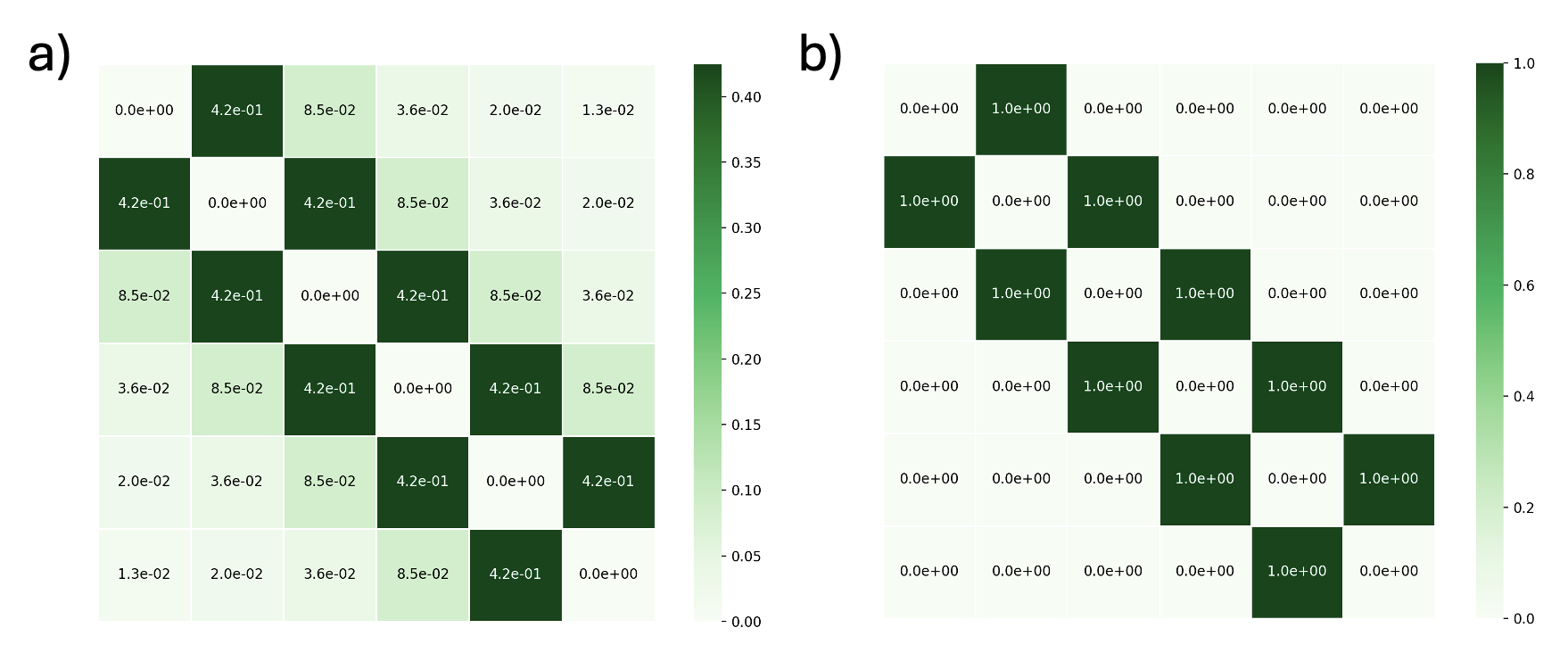}
    \caption{Matrix representation of the (a) classical Laplacian and (b) fractional Laplacian with $s=0.5$ b).}
    \label{fig:matrix}
\end{figure}

The nonlocal interaction modeled by the discrete Fractional Laplacian can be visualized with a matrix representation. In this context, the matrix form refers to the discretized operator acting on a vectorized field, where each matrix entry represents the strength of interaction between two points. For the classical Laplacian, the resulting matrix is sparse, with nonzero elements only for nearest neighbors (i.e., a tridiagonal structure), as shown in Figure \ref{fig:matrix}. In contrast, the matrix representation of the fractional Laplacian (here for $s=0.5$) Figure \ref{fig:matrix}  b), contains nonzero off diagonal terms for all matrix elements, which represents the nonlocal nature of the interaction. It can also be seen that the off diagonal terms of the fractional Laplacian matrix rapidly fall off with nonlocal range $m$.

 In the spectral approach, the Hamiltonian in equation \ref{eq:H} iteratively operates on an initial basis vector $\delta_0$, which propagates the energy in space. The sequence of energy states obtained after $N$ iterations of the Hamiltonian is given by

\begin{equation}
    \{ \delta_0,H_{s,\epsilon}^1\delta_0,H_{s,\epsilon}^2 \delta_0,…,H_{s,\epsilon}^N \delta_0 \} \to \{H_{s,\epsilon}^k \delta_0 \}_{k=0}^N .
\end{equation}

A Grahm-Schmidt orthogonalization (without normalization) is performed on this sequence to obtain a new sequence 

\begin{equation}
    \{ \varphi'_0,\varphi'_1,\varphi'_3,…,\varphi'_N\} \to \{\varphi'_k \}_{k=0}^N .
\end{equation}

Each of the vectors $\varphi'_k$ represents the new information obtained by the $k^{th}$ iteration of the Hamiltonian. The orthogonalization of the sequence also allows the definition of a mathematical distance equation in the Hilbert space (i.e., the generalization of the Euclidean vector distance to infinite-dimensional space). For any vector in the Hilbert space $\nu$, such that $\nu\neq\delta_0$, the mathematical distance between $\nu$ and the $k^{th}$ element of the orthogonalized sequence $\{\varphi'_k \}_{k=0}^N$ is given by

\begin{equation}
D_{s,c}^N \equiv \sqrt{1-\sum_{k=0}^N \left(\frac{\langle \nu,\varphi_k'\rangle}{\|\nu\|\|\varphi_k'\|}\right)^2}
\label{eq:dist}.
\end{equation}

 This distance in Hilbert space can be interpreted as the probability of transport away from the initial state. Mathematically, the extended states conjecture states that $\lim_{N \to \infty}D_{s,\epsilon}^N\neq 0$ implies the existence of a continuous spectrum for the Hamiltonian, i.e., the existence of extended states. The opposite statement, $\lim_{N \to \infty}D_{s,\epsilon}^N = 0$, implying, increased probability for localization, is less rigorous since, in the absence of extended states, the remaining spectrum can consist of both a singular part (i.e., eigenvalues) and a poorly-behaved singular-continuous part. The implementation of Equation \ref{eq:dist} we call the Fractional Laplacian Spectral Method (FLSM).

 \vspace{3mm}

 In \autoref{sec:Analysis}, we will use the extended states conjecture to calculate the probability for the existence of extended states for Hamiltonian operators that correspond to the nonlocality and stochasticity measured in each PK-4 pressure-current dataset. The changes in the calculated spectra will be used to understand changes in the stability and dynamics of the observed global structural states of the dusty plasma clouds. In addition, we will investigate how high probability for extended states at characteristic scales in Hilbert space can be linked to the characteristic scales of dust particle jumps observed in the PK-4 experiments. The dust particles in these experiments are mostly confined within self-organized structures reminiscent of liquid-crystalline filamentary states. However, examination of the video data reveals that, occasionally, individual dust particles can escape from the lattice-like filamentary structures and perform big jumps. These are similar to L\'{e}vy flights, which may quickly take them outside the field of view, however these are not scale-free and infinite, but can be quantified. Thus, while they are larger than the typical displacements in the system they are not outside of the scale of the system and therefore are somewhere between L\'{e}vy flights and the mean free path. These are of the proposed scale $\lambda_{mfp}\ll\lambda_q\ll \textit{L\'evy}$. This can be visually observed in raw video data of PK-4 but in \ref{sec:scaling} we present quantified evidence of this from particle tracking data as well.

 \vspace{3mm}
 
 The theoretical framework described so far allows for a connection between the experimentally-extracted nonextensive statistical parameter $q$ and the fraction $s$ on the Laplacian kinetic term of the Hamiltonian. In the previous study \cite{Andrew2025} we calculated the MSD $\alpha$ and $q_p$ and now we show the converted $s_\alpha$ and $s_q$.

 \begin{table}[H]
    \centering
    \small
    \renewcommand{\arraystretch}{1.5}
    \begin{tabular}{|c|c|c|c|c|c|c|c|c|c|c|c|}
        \hline
        \multicolumn{2}{|c|}{Pressure–Current} & $\alpha_{\parallel}$ & $\alpha_{\perp}$ & $q_{p\parallel}$ & $q_{p\perp1}$ & $q_{p\perp2}$ & $s_{\alpha\parallel}$ & $s_{\alpha\perp}$ & $s_{q\parallel}$ & $s_{q\perp1}$ & $s_{q\perp2}$ \\
        \hline
        \multirow{3}{*}{70 Pa} & 1 mA    & 2.01 & 1.05 & 1.81 & 1.40 & 1.60 & 0.498 & 0.952 & 0.482 & 0.800 & 0.700 \\ 
        \cline{2-12}
                               & 0.7 mA  & 1.68 & 1.13 & 1.70 & 0.98 & 1.25 & 0.595 & 0.885 & 0.647 & 1.010 & 0.875 \\ 
        \cline{2-12}
                               & 0.35 mA & 2.05 & 1.45 & 1.55 & 0.96 & 1.38 & 0.488 & 0.690 & 0.725 & 1.020 & 0.810 \\ 
        \specialrule{1.5pt}{0pt}{0pt}
        \multirow{3}{*}{46 Pa} & 1 mA    & 2.40 & 0.925 & 1.44 & 0.98 & 1.27 & 0.417 & 1.081 & 0.780 & 1.010 & 0.865 \\ 
        \cline{2-12}
                               & 0.7 mA  & 1.54 & 1.10  & 1.41 & 0.95 & 1.20 & 0.649 & 0.909 & 0.795 & 1.025 & 0.900 \\ 
        \cline{2-12}
                               & 0.35 mA & 2.03 & 1.40  & 1.47 & 0.95 & 1.21 & 0.493 & 0.714 & 0.765 & 1.025 & 0.895 \\ 
        \specialrule{1.5pt}{0pt}{0pt}
        \multirow{3}{*}{30 Pa} & 1 mA    & 1.68 & 1.18  & 1.33 & 0.99 & 1.22 & 0.595 & 0.847 & 0.835 & 1.005 & 0.890 \\ 
        \cline{2-12}
                               & 0.7 mA  & 1.50 & 1.04  & 1.25 & 0.94 & 1.28 & 0.667 & 0.962 & 0.875 & 1.030 & 0.860 \\ 
        \cline{2-12}
                               & 0.35 mA & 1.67 & 0.88  & 1.34 & 1.07 & 1.10 & 0.599 & 1.136 & 0.830 & 0.965 & 0.950 \\ 
        \hline
    \end{tabular}
    \caption{Fitted and calculated parameters for the 9 pressure-current cases in the PK-4 experiments. The values of $\alpha$ and $q$ were extracted from fits to MSD plots and position histograms, respectively. The values of $s_{\alpha}$ were calculated using the scaling relation $s_\alpha = 1/\alpha$. The values of $s_q$ were extracted using the scaling relation $s_q = \frac{3 - q}{2}$ if $q < \frac{5}{3}$ and $s_q = \frac{3 - q}{2q-2}$ if $q \geq \frac{5}{3}$.}
    \label{tb:sq_table}
\end{table}

This provides a parameter space $0.4<s<1.2$ for the fractional Laplacian in the spectral calculations.  The next step is to quantify the dimensionless disorder for each of the PK-4 pressure-current cases, which will provide the key input for the potential energy term of the Hamiltonian.

\section{Disorder Parameter}
\label{sec:Disorder}

To quantify the structural disorder for each dusty plasma cloud over time, we combined two complementary measures of disorder: a density-based disorder and an orientational disorder based on bond angles. The particle positions were obtained from video data using particle tracking techniques and assuming camera pixel size of $dx = 14.20~\mu\text{m}$. This allowed for all spatial measurements, including sampling radii and inter-particle distances, to be computed in meaningful physical units. The density-based disorder was computed by sampling 60 randomly positioned circular windows around each particle. The center of each window was uniformly drawn from within a radial distance of $30dx$ (30 pixels) from the particle, and each window had a fixed radius of 30 pixels. This window radius was chosen to be sufficiently large to capture mesoscale fluctuations in particle concentration while remaining much smaller relative to the scale of the cloud, which extends over about 1000 pixels in the horizontal direction and about 150 pixels in the vertical direction. Within each window, we counted the number of neighboring particles and calculated the variance across all samples. This variance quantifies the local density inhomogeneity around each particle, with higher values indicating greater spatial disorder due to clustering or irregular voids. Formally, the density-based disorder for particle $i$ is expressed as

\begin{equation}
    c_i^{(\text{density})} = \operatorname{Var}(N_1^{(i)}, N_2^{(i)}, \dots, N_{60}^{(i)}),
\end{equation}
where $N_j^{(i)}$ is the number of neighbors found in the $j$-th sample window around particle $i$. 

\vspace{3mm}

Next, we evaluated the angular distribution of neighbors around each particle using the bond orientational order parameter $G_6$. In all cases, the filamentary structures in the dust clouds seemed to exhibit hexagonal alignment within the 2D plane of the illumination laser sheet, which is why $G_6$ was selected as appropriate orientational measure. For a given particle, we identified its six nearest neighbors using a KD-tree search and computed the angle $\theta_{ij}$ between the particle and each neighbor $j$. These angles were used to define the complex order parameter,
\begin{equation}
    G_6(i) = \frac{1}{6} \sum_{j=1}^{6} e^{i6\theta_{ij}},
\end{equation}
which approaches a magnitude of 1 for a perfectly hexagonal (triangular lattice) configuration. We then defined the structure-based disorder as the deviation from perfect orientational order:
\begin{equation}
    c_i^{(\text{structure})} = 1 - |G_6(i)|.
\end{equation}
The two disorder metrics were combined linearly using a tunable mixing parameter $\beta$, 

\begin{equation}
    c_i = \beta c_i^{(\text{density})} + (1 - \beta) c_i^{(\text{structure})}
\end{equation}
with weight ($\beta = 0.6$) used in our analysis. We weight the density variation slightly more because of the filamentary structure of the PK-4 clouds where interparticle spacing and hexagonal alignment fluctuates with time.
\vspace{3mm}

\begin{figure}[H]
    \centering
    \includegraphics[width=100mm]{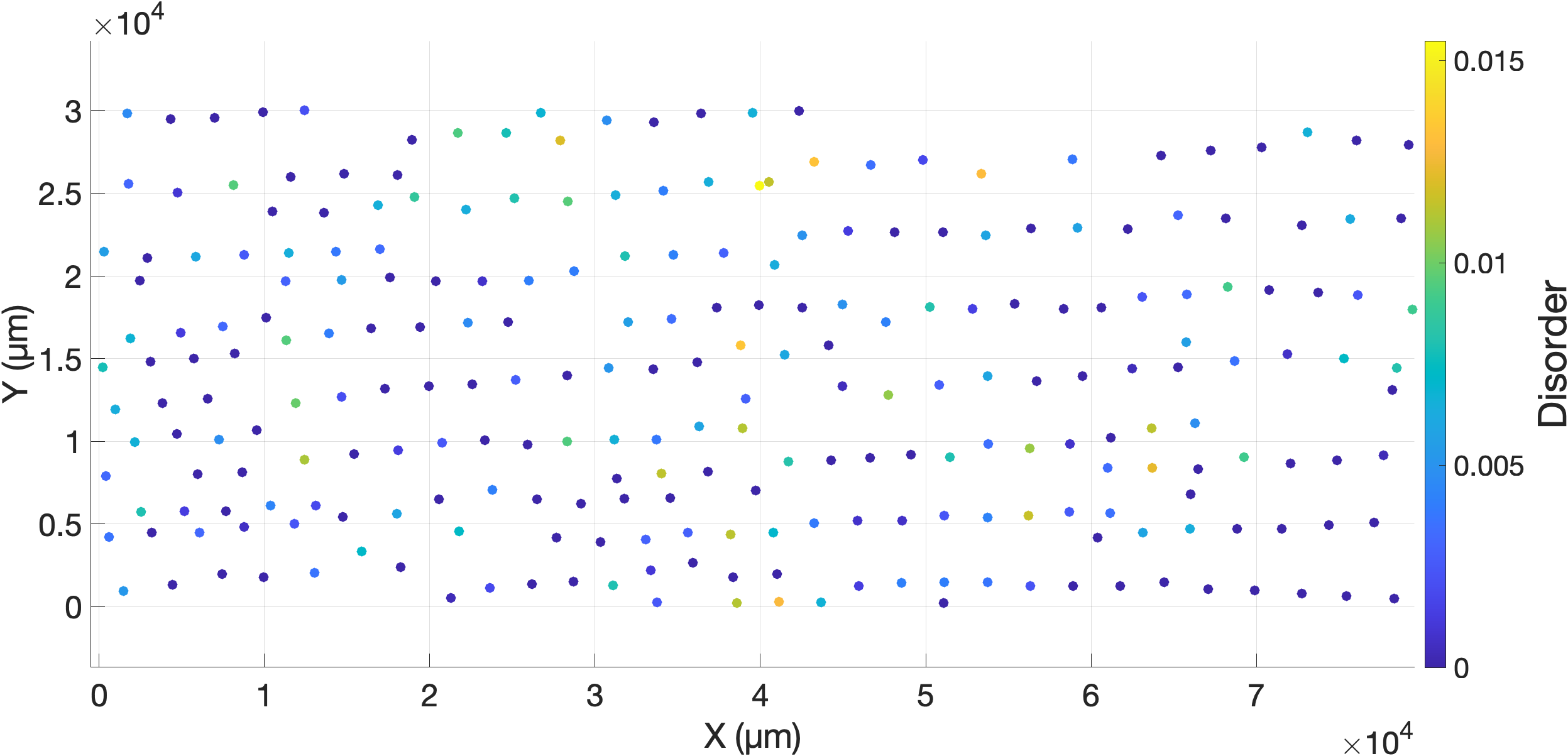}
    \caption{Disorder visualization at a single frame for the case of 70 Pa and 0.35 mA. Particle positions are color-coded by normalized disorder values.}
    \label{fig:disorder_single_frame}
\end{figure}

\begin{figure}[H]
    \centering
    \includegraphics[width=100mm]{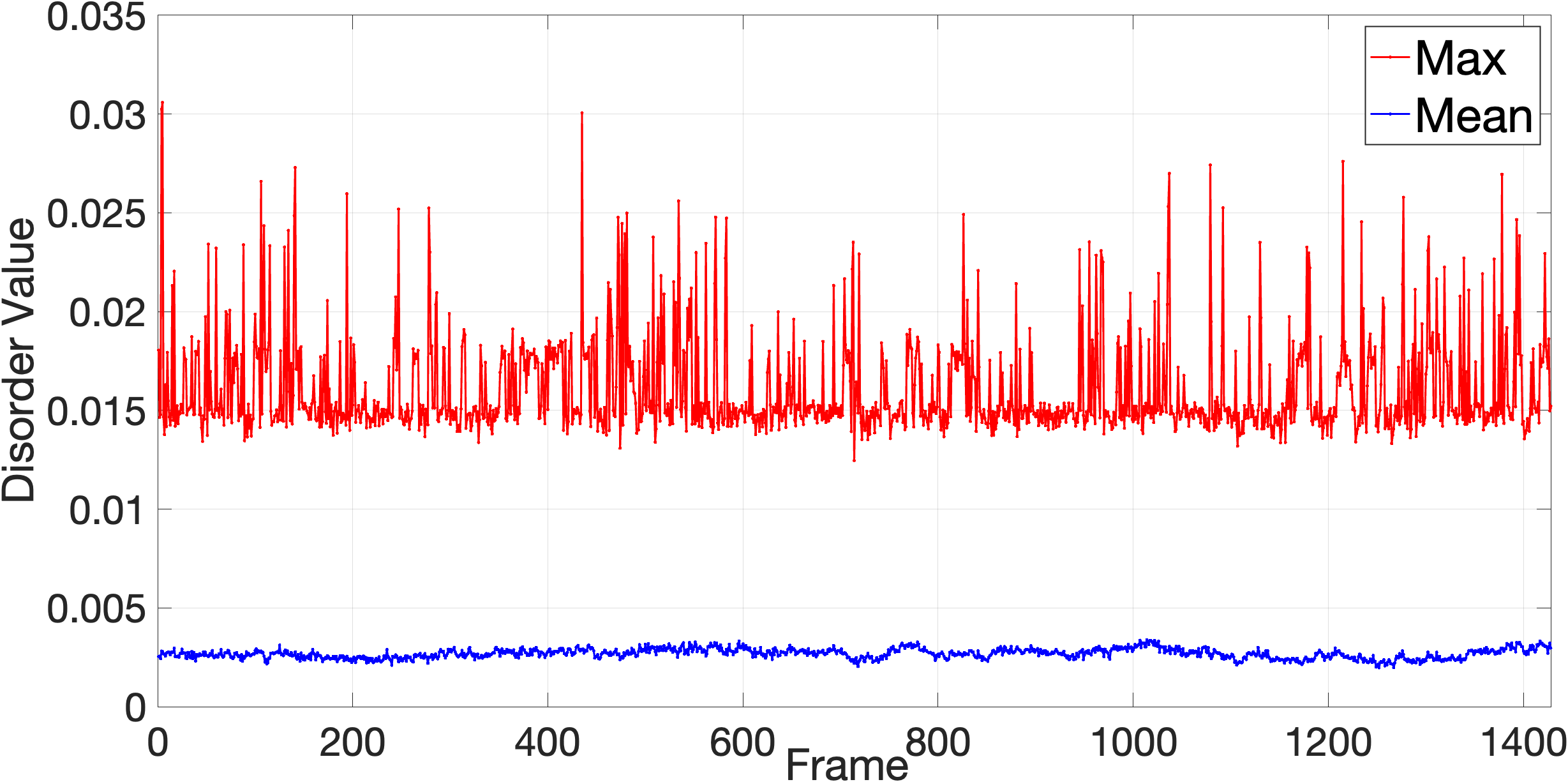}
    \caption{Temporal evolution of disorder across all frames for the case of 70 Pa and 0.35 mA. The plot shows the mean and maximum disorder per frame.}
    \label{fig:disorder_over_time}
\end{figure}

To ensure comparability across frames and across experiments, we performed a global normalization of the combined disorder metric using fixed reference values. The minimum disorder value was obtained from a simulated hexagonal lattice with inter-particle spacing set to match the experimental average. The maximum disorder value was computed from a uniform random distribution of the same number of particles, generated within the same field of view. The final normalized disorder metric for each particle was defined as

\begin{equation}
    c_i^{(\text{norm})} = \frac{c_i - c_{\min}}{c_{\max} - c_{\min}},
\end{equation}

which maps values to the interval $[0, 1]$, where 0 corresponds to perfect crystalline order and 1 to maximal disorder. To address occasional numerical instabilities, such as extremely large variance values due to sparsely populated regions or edge effects, we applied a correction step. Any disorder value exceeding a threshold (e.g., 0.5) was flagged as an outlier and replaced with a corrected value defined as 1.25 times the second-highest valid disorder in that frame. This ensured that rare pathological points did not distort color scales or skew statistical averages. All disorder values were visualized over time using color-mapped scatter plots with a consistent turbo colormap and fixed scaling. A representative scatter plot is shown in Figure \ref{fig:disorder_single_frame}. Additionally, the temporal evolution of the minimum, mean, and maximum disorder across all frames was tracked to observe changes in the system's spatial organization, as shown in Figure \ref{fig:disorder_over_time}. Overall, the tunable parameters were chosen so that the disorder map of the dust particle positions aligns with what is expected from observation. For example, particles in straight filaments should have a low disorder $\sim 10^{-5}$ while particles not aligned in filaments or particles that are visibly crowded together should have a higher disorder $\sim 10^{-1}$.

From the plot we see that the maximum disorder fluctuates much more than the mean disorder. This provides some information about the system dynamics suggesting that dust particles in disordered regions participate in large structural changes from one frame to the next. 

\begin{figure}[H]
    \centering    \includegraphics[width=140mm]{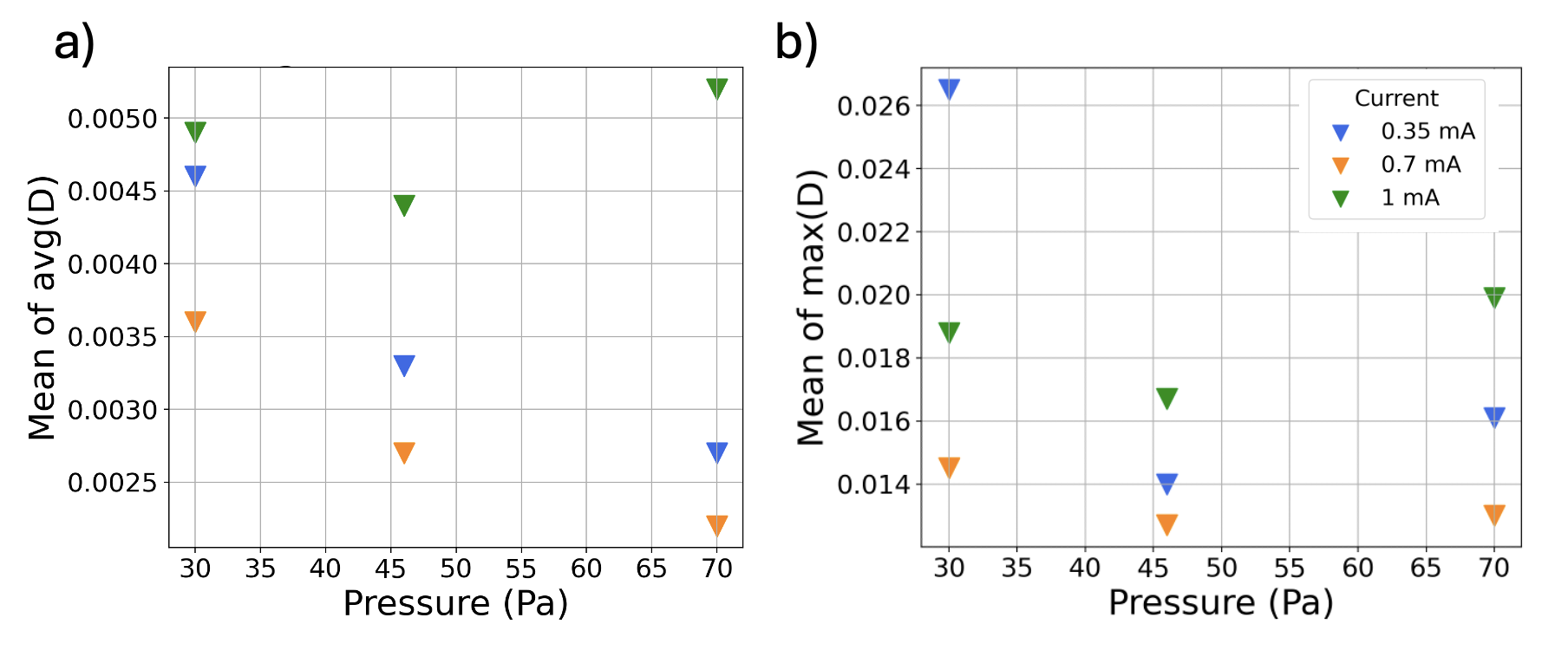}
    \caption{Disorder for all nine pressure and current conditions: (a) mean of the average disorder for all frames, and (b) mean of the max disorder for all frames.}
    \label{fig:disordervpressure}
\end{figure}

We see that for most cases the mean disorder decreases with increasing pressure. This is also in agreement with Part One of this study \cite{Andrew2025} as the dust kinetic temperature was also found to decrease with increasing pressure. However, the maximum disorder shows a nonlinear relation with pressure, with minimum values at 46 Pa. As discussed in Part One, the cloud in the 46 Pa cases was larger, which results in higher dust densities. For that cloud, it can be expected that the outer layers of dust particles provide stronger confinement of particles in the interior, thus improving stability. The higher fluctuation of the maximum disorder in Figure \ref{fig:disorder_over_time} is also in agreement with Part One of this work, where we discussed that, while the majority of the dust particles are in a stable configuration, there exists a subpopulation of particles with much higher energy as evident from the 'fat' tailed velocity distributions. These higher energy dust particles make big jumps across the cloud, which contribute to the maximum disorder over time.

\section{Spatial Scaling}
\label{sec:scaling}

In the previous sections, we discussed how to obtain the fraction $s$ on the Laplacian and the dimensionless disorder $c$, which are the two main inputs in the Fractional Laplacian Spectral Method (FLSM) \ref{eq:dist}. As mentioned previously, FLSM calculates whether a continuous spectrum for a given Hamiltonian exists, which corresponds to the existence of extended energy states for the system described by that Hamiltonian. The FLSM calculation can be performed with respect to any reference vector $\nu$ in Hilbert space. The vector $\nu$  is a linear combination of the standard basis vectors $\delta_i$ from Eq. \ref{eq:H}, and the number of basis functions can be varied. In other words, $\nu$ is simply a subspace of the Hilbert space and increasing the number of $\delta_i$ basis vectors used to define $\nu$ increases the size of the subspace. Mathematically, if continuous spectrum exists, performing the FLSM calculation with respect to subspaces of increasing size should always yield a nonzero value of the distance parameter in Eq. \ref{eq:dist}. Physically, if the distance parameter limits to a non-zero value with respect to some subspaces, but limits to zero at other scales, the system described by the Hamiltonian is expected to exhibit characteristic length scales of transport  defined by the combined effects of nonlocality and disorder. In the PK-4 experiments, the characteristic scale of diffusive transport can be understood from the characteristic step-size displacement of the dust particle diffusion.

Historically, studying diffusion was performed in part by analyzing step-size displacement of the particles. This provides spatial information important to the dynamics of the system. For example, the microscopic dynamics are certainly going to be different for a system containing much larger scale displacements, like L\'{e}vy flights. Mathematically, L\'{e}vy flights are characterized as being scale-free and thus infinite in size. That is to say, there is not a rigorous observational way to define if a displacement is a L\'{e}vy flight based on a simple criterion such as, for example, the displacement being found to be ten times the average displacement or greater. However, in our experiments, the particles (and their jumps) are tracked at the kinetic level, which allows for some quantification of the characteristic step-size of particle diffusion and the identification of large jumps, even if they are not L\'{e}vy flights. Here we take a large portion of the dust particle tracks for each pressure-current case and find the average step size, the maximum step size, and a percentage of how many of the step sizes were ten times greater than the average.

\begin{figure}[H]
    \centering    \includegraphics[width=140mm]{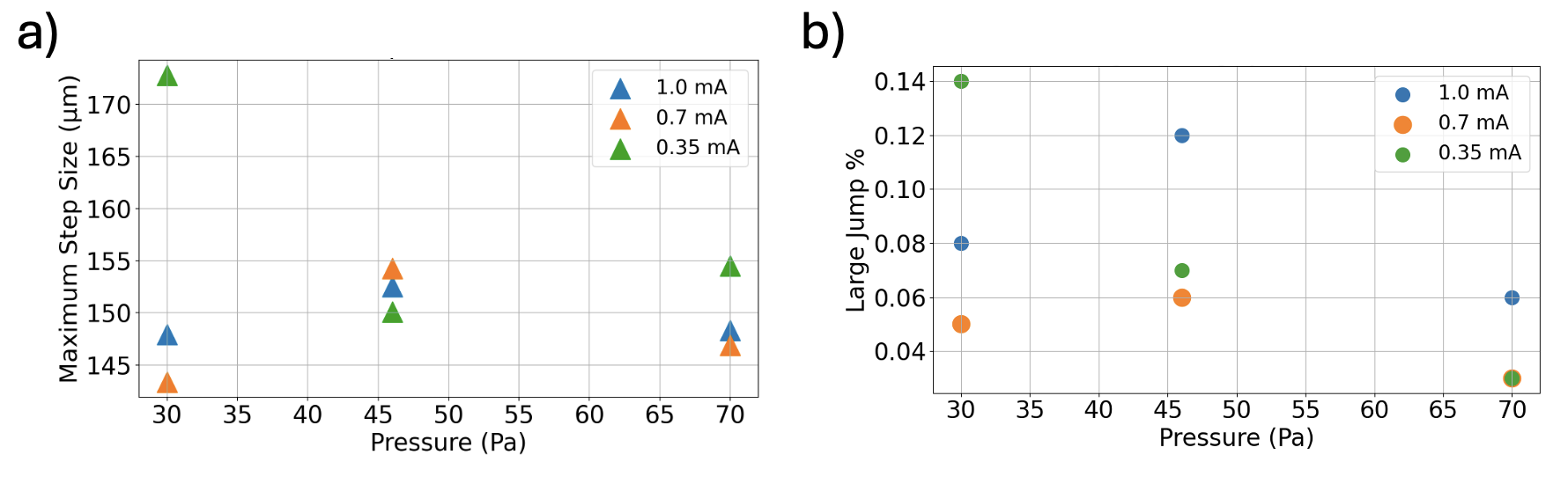}
    \caption{a) Maximum step-size for all pressure-current cases. b) Percentage of particle displacements that are ten times the average step-size for all pressure-current cases.}
    \label{fig:scaling}
\end{figure}

We found that, after drift motion was carefully subtracted from the dust trajectories, the average displacement in between two successive frames (frame rate $70.1 fps$, yielding $\Delta t=0.014 s$) across the pressure-current cases was approximately one pixel or $14 \mu m$. This is presumably due to background collisions or local electrostatic fluctuations. Due to both spatial and time resolution in these experiments, this is therefore the smallest spatial scale to observe. Figure \ref{fig:scaling} a) shows that the maximum particle displacements are slightly larger than ten times the average particle displacements. These are characterized as "large jumps" for the remainder of the paper. The percentage of "large jumps" in the system, as shown in Figure \ref{fig:scaling} b), is in the range 0.03\% - 0.14\% for all pressure-current cases, with overall higher numbers occurring at lower pressures. With around 440 particles in a frame and 1400 frames of data, this yields approximately 616,000 total displacements of which around 500 are "large jumps". Note that the video data clearly showed frequent instances of particles that make much larger jumps, on the order of $1 mm$ or bigger, which resembles L{\'e}vy flights. However, those were omitted from the analysis due to difficulties in obtaining accurate tracks with the particle tracking software.

When breaking down the different scales of step-sizes (mean, large, and clustered), we saw that the $\parallel$-step-sizes tend to decrease more with increasing pressure than the $\perp$-step-sizes. This trend was strongest for the mean step-size (Pearson $r = -0.901$ for $\parallel$ vs. pressure, compared to $r = -0.765$ for $\perp$ vs.  pressure). The trend persisted across the large and clustered steps, though with decreasing magnitude and statistical significance. We also assessed whether large jumps exhibited a distinct directional bias by comparing the angular distributions of mean displacements and "large jump" displacements. The average angular orientation with respect to the direction of the electric field was \(1.1^\circ\) for mean displacements and \(7.5^\circ\) for large displacements. A two-sample t-test indicated a trend toward significance \(p=0.0837\), where the p-value reflects the probability of observing this result if no true difference exists, suggesting a modest directional shift. To verify this without assuming normality, we performed a Mann--Whitney U test, which yielded a comparable result (\(p = 0.0729\)). These results suggest that large jumps may be preferentially oriented cross field angles than typical displacements, though the evidence remains suggestive rather than conclusive. We also found that successive "large jump"  occasionally take place. This is a scale larger than $150 \mu m$ of dust displacements, where dust displacements on the order of the interparticle separation happen directly after a previous "large jump". This can be seen in the below plot which gives the percentage of "large jumps" that recently (within 10 frames) had another "large jump" happen. We call this a "cluster of large jumps", where "large jumps" happen relatively soon after another and could span a larger distance. This distance is roughly twice as large since the jump clusters are typically pairs of "large jumps". However, these clusters of large jumps have a wide distribution of angles. This means the "clusters of large jumps" are not found to have a directional bias. 

\begin{figure}[H]
    \centering    \includegraphics[width=80mm]{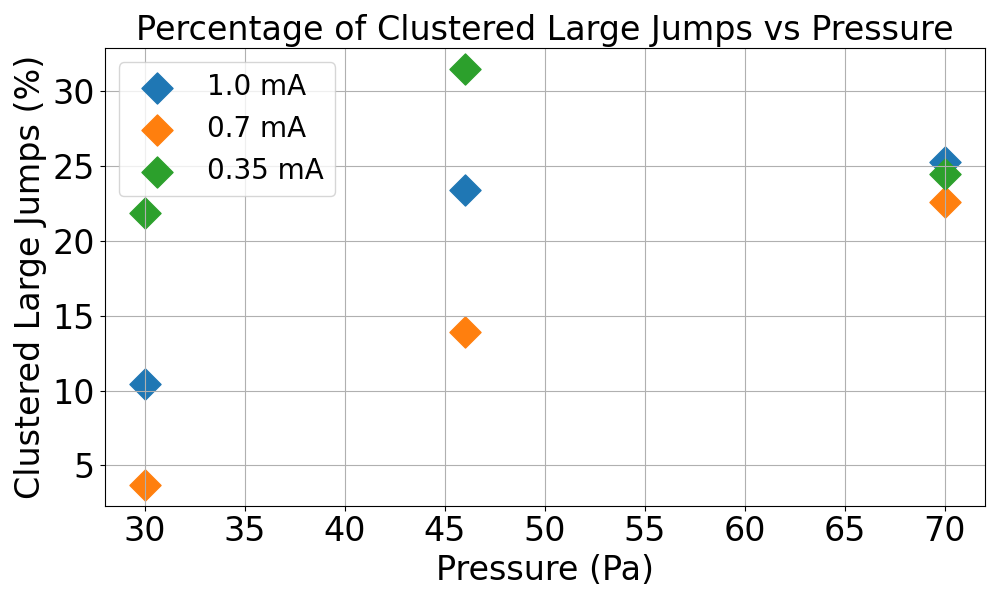}
    \caption{Percentage of "large jumps" that are followed by another "large jump" within 10 frames.}    
    \label{fig:cluster}
\end{figure}

Among all pressure–current combinations analyzed, a total of 19 clusters of large particle displacements were identified across the pressure-current datasets. Of these, only three clusters exhibited a directional standard deviation less than $30^\circ$, indicating a relatively coherent directional alignment. These three "clusters of large jumps" were aligned in the cross field direction, thus leading to a total displacement in the cross field direction much larger than any other displacement around $250 \mu m$. Specifically, two of these clusters originated from the $30~\mathrm{Pa}$ and $0.35~\mathrm{mA}$ condition, while the third occurred under $70~\mathrm{Pa}$ and $0.70~\mathrm{mA}$.

\section{FLSM Results and Analysis}
\label{sec:Analysis}

In this section, we present the results from solving Equation \ref{eq:dist} for Hamiltonian operators encompassing a wide parameter space of nonlocality and stochasticity conditions, which includes the ranges of parameters extracted from the PK-4 experiments. The input parameters for the model are: the fraction on the Laplacian $s$ (which quantifies the type and strength of nonlocality), the number $m$ of terms in the series representation of the fractional Laplacian (which quantifies the range of nonlocality), the dimensionless disorder $c$ (which quantifies fluctuations in potential energy), the number of trials to average over different random realizations of the dimensionless disorder (here we did 8 trials for every parameter combination), the reference scale $\nu$ (which is a subspace in Hilbert space), and the number of time-steps $N$ (which corresponds to the number of iterative applications of the Hamiltonian). The series representation of the fractional Laplacian is valid for infinite number of terms $m$ in the series \cite{padgett_anomalous_2020}, while the extended states conjecture \cite{liaw_approach_2013} is true for infinite number of time steps $N$. While this is computationally not feasible, this limitation can be overcome by reasonable truncation of the fractional Laplacian series and running the simulation for sufficient number of time-steps and either extrapolating the distance value at infinity from fitting or performing asymptotic analysis of the final time-steps.  

\begin{figure}[H]
    \centering    \includegraphics[width=160mm]{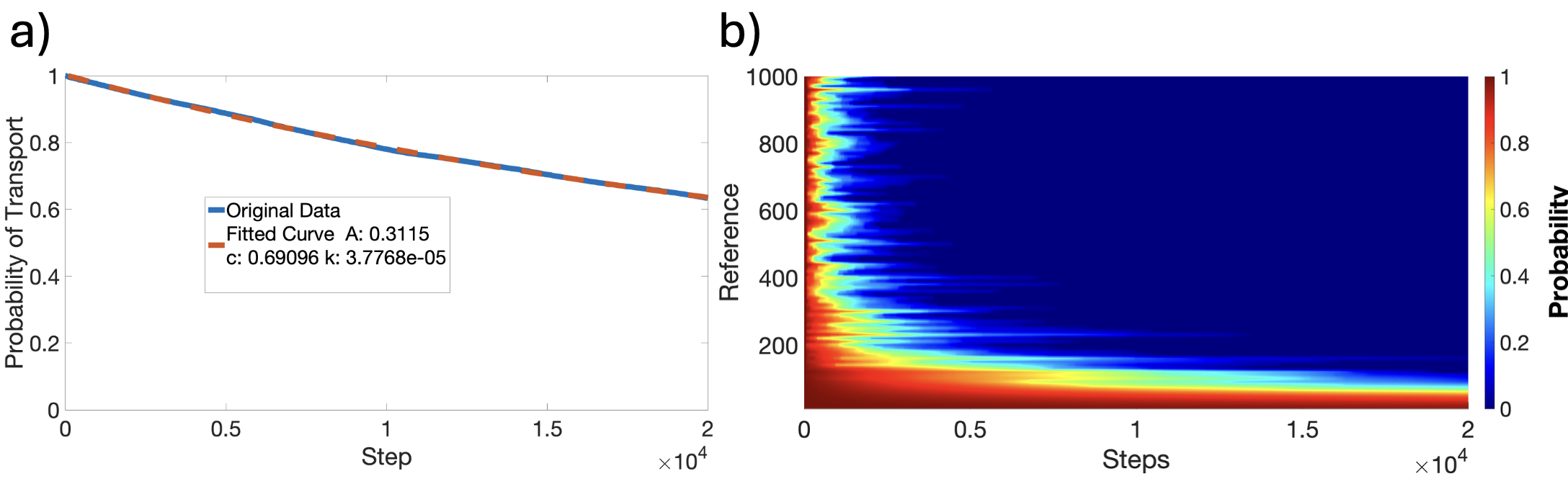}
    \caption{Probability for transport $D_{s,c}$ for parameters $s=0.65$, $c=0.01$, $m=150$, and $N=20000$ time-steps. a)  $D_{s,c}$ as a function of timestep for reference scale $\nu=60$, along with equation fit. b) The value of $D_{s,c}$ as a function of time-step calculated for reference scales from $\nu=10$ to $\nu=1000$ in increments of $\Delta\nu=10$.}
    \label{fig:example}
\end{figure}

Since the off-diagonal terms of the fractional Laplacian fall off rapidly, we choose a cutoff value of $m=75$ to account for sufficient nonlocal range behavior. For this number of terms in the series, only a small remainder is discarded from the calculation, while maintaining computational efficiency. The effect of different series truncations is discussed in more detail in \cite{padgett_anomalous_2020} and shown in Table 1 of that paper. Next, since we cannot perform infinite iterations of the Hamiltonian, we perform as many as computationally feasible and then analyze the asymptotic behavior. Curve fitting to the limiting behavior was implemented for a few parameter combinations. Figure \ref{fig:example} shows a plot of the distance $D^N_{s,c}$ for the specific case $s=0.65$, and $c=0.01$. We tested 6 different curve fitting relations to help find the relation most closely corresponding to the limiting behavior of $D_{s,c}$ over many simulation time-steps. The form $D=A+be^{-k t}$ consistently yields the least error over multiple parameter tests. Since the best fits correspond to an exponential decay function, this provides confidence that the chosen number of time-steps for the simulation is sufficiently large to reveal the limiting behavior of the distance parameter. 

The value of $D_{s,c}$ as a function of timestep for a single combination of parameters is shown in Figure \ref{fig:example} a). A 3D plot, for the same set of parameters over many reference scales is shown in Figure \ref{fig:example} b). These figures are useful to see how the simulation progresses over time and visualize numerical fluctuations (e.g., due to the random realization of disorder). The fits to specific combinations of parameters are useful to infer the functional form of the asymptotic behavior, here - an exponential decay towards a constant value. However, we find that analyzing the limiting value at infinity for a large-scale parameter scan can be most efficiently performed by asymptotic analysis of the final time-steps. To determine whether $D_{s,c}$ may indicate probability of transport, we look at the asymptotic saturation by analyzing the tail behavior of $D_{s,c}$ after $N=8000$ time-steps were extracted from each parameter combination. This was done by computing the mean of the absolute rolling slope over the final 200 time-steps of each trajectory

\begin{equation}
    \bar{\sigma} = \frac{1}{k - 1} \sum_{i=1}^{k-1} \left| \frac{D(N_{i+1}) - D(N_i)}{\Delta N} \right|.
\end{equation}

\begin{figure}[H]
    \centering    \includegraphics[width=120mm]{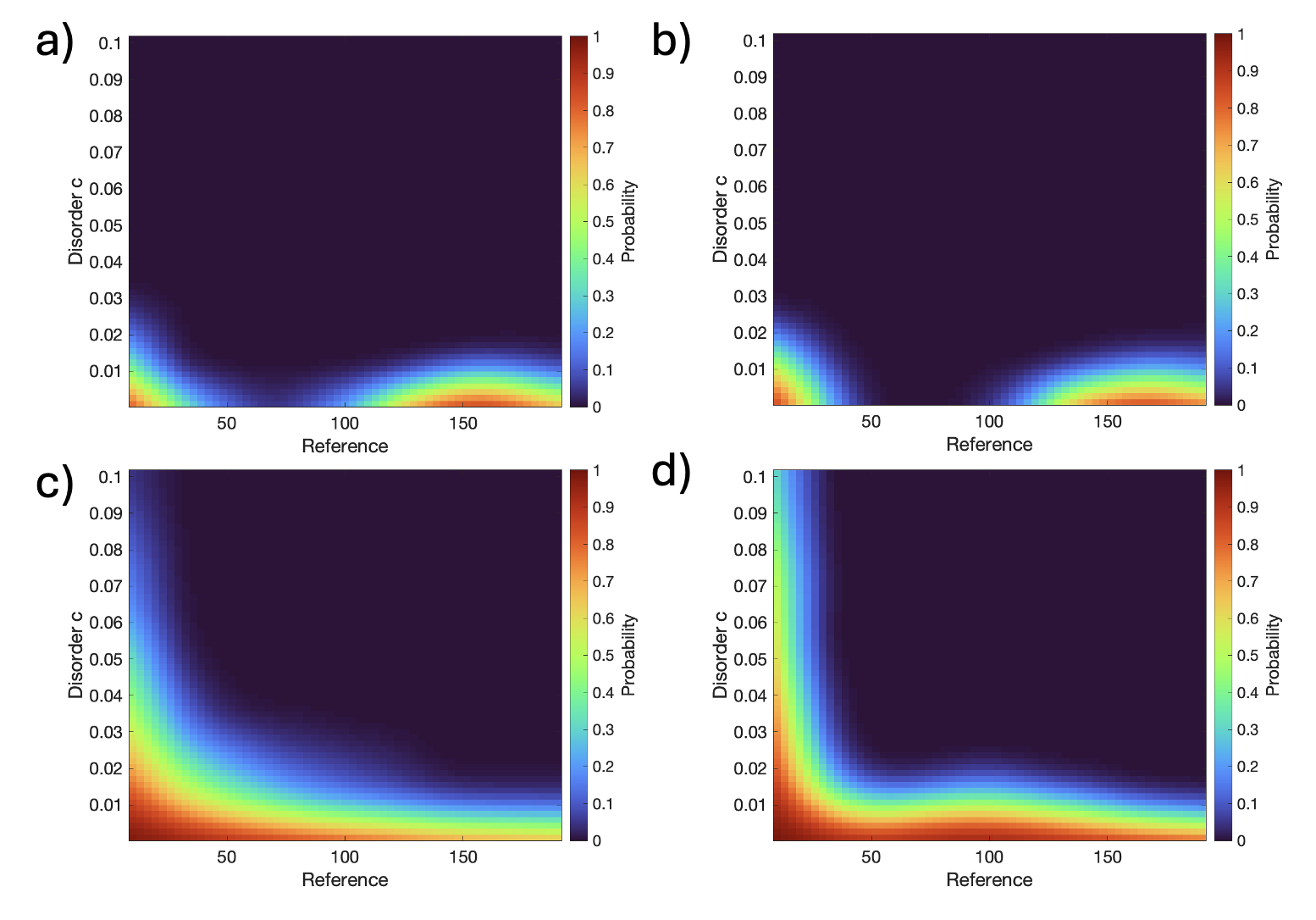}
    \caption{Probability of transport for fixed  values of $s$: a) $s=0.5$ (L\'{e}vy process) b) $s=0.8$ (superdiffusion) c) $s=1.0$ (classical diffusion) d) $s=1.3$ (subdiffusion)}    
    \label{fig:cvr_all}
\end{figure}

\begin{figure}[H]
    \centering    \includegraphics[width=160mm]{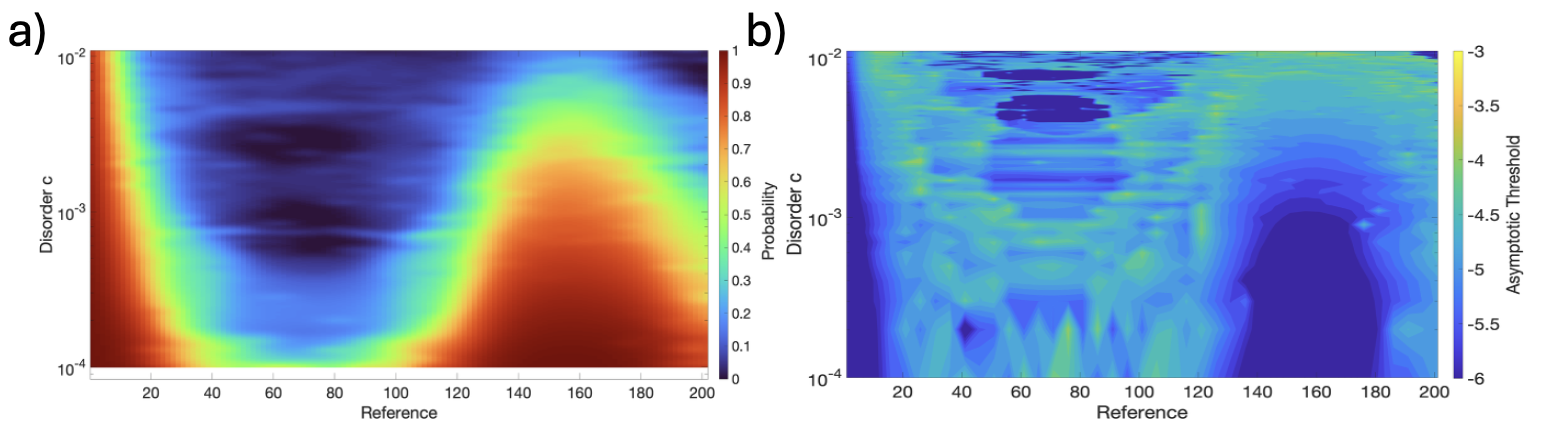}
    \caption{(a) Probability for transport for fixed parameter $s=0.66$, corresponding to the crossover between superdiffusion and a L{\'e}vy process. (b) Corresponding asymptotic convergence map. Darker blue regions indicate very strong convergence.}    
    \label{fig:cvr_066}
\end{figure}

\begin{figure}[H]
    \centering    \includegraphics[width=120mm]{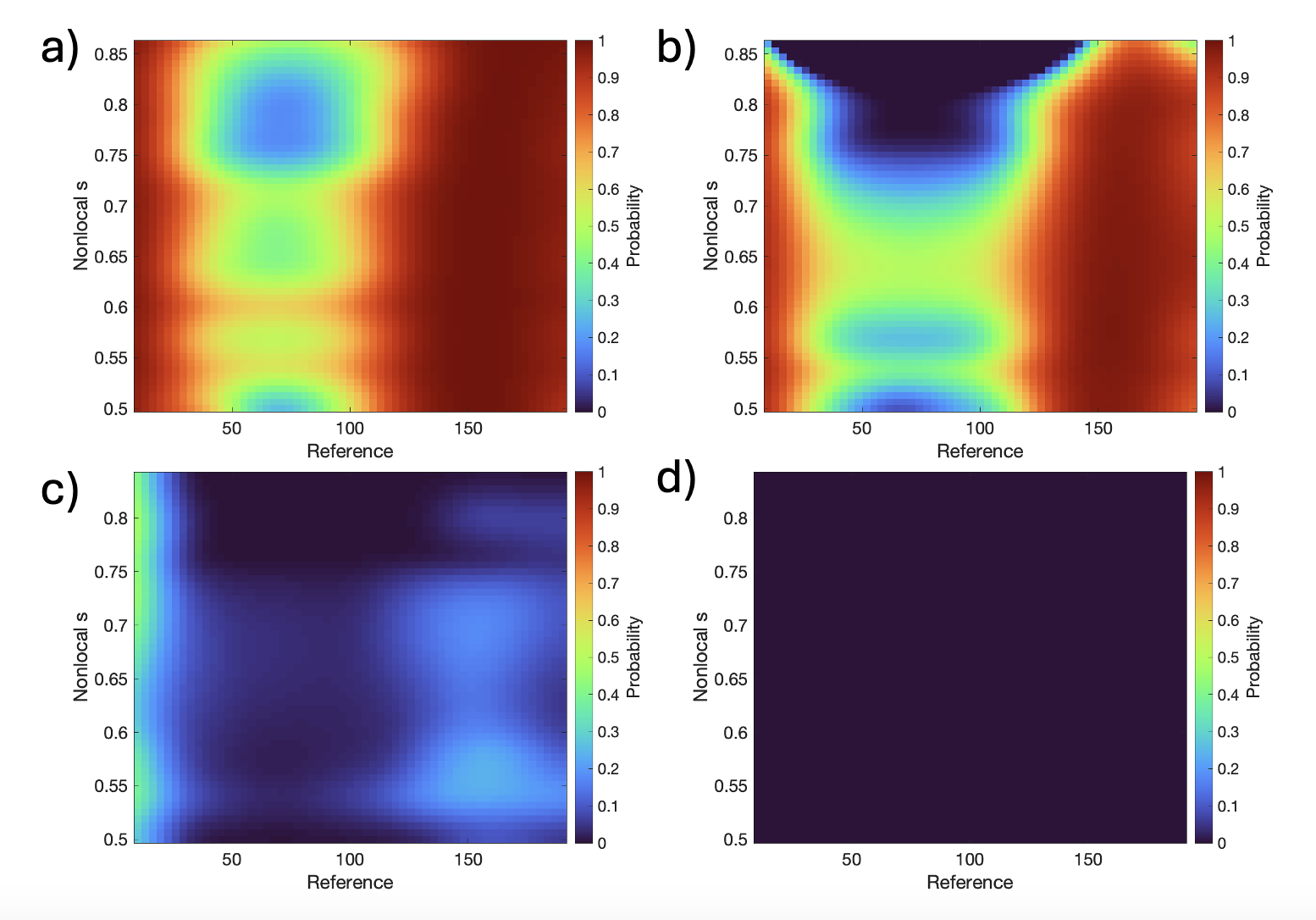}
    \caption{Probability of transport for fixed values of disorder a) $c=0.0001$, b) $c=0.001$, c) $c=0.01$, and d) $c=0.1$. }    
    \label{fig:svr_all}
\end{figure}

\begin{figure}[H]
    \centering    \includegraphics[width=140mm]{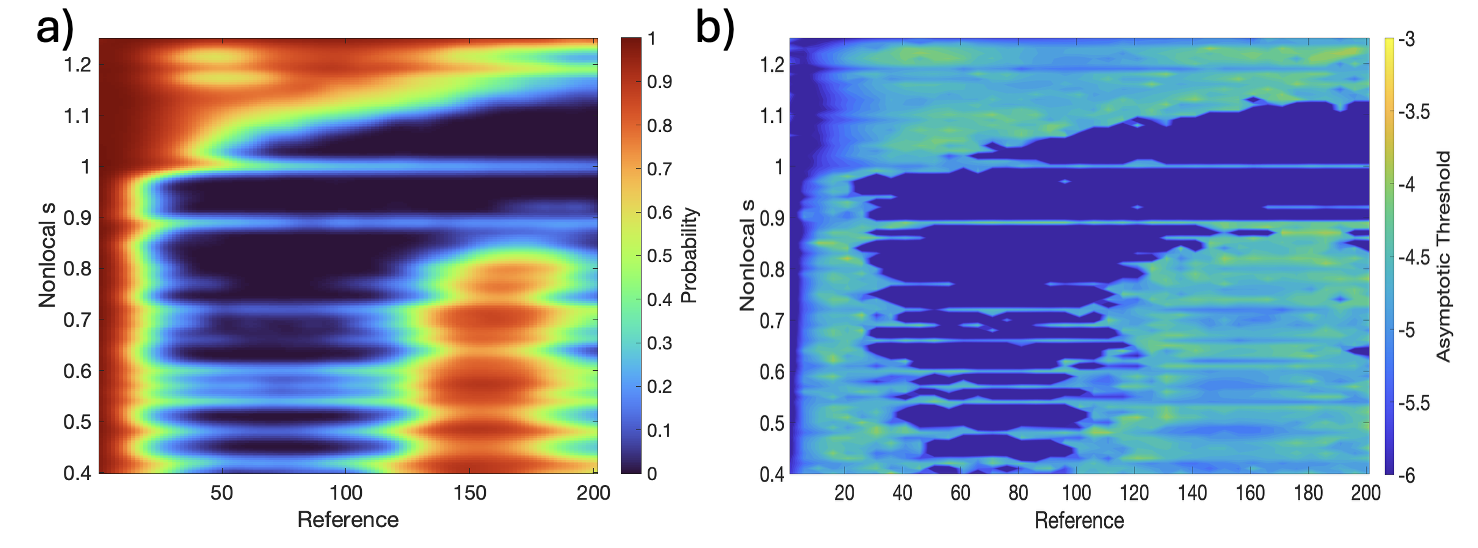}
    \caption{ (a) Probability of transport for fixed disorder $c=0.0035$. (b) Corresponding asymptotic convergence map with darker blue indicating very strong convergence.}    
    \label{fig:svr_c0_0035}
\end{figure}

\begin{figure}[H]
    \centering    \includegraphics[width=160mm]{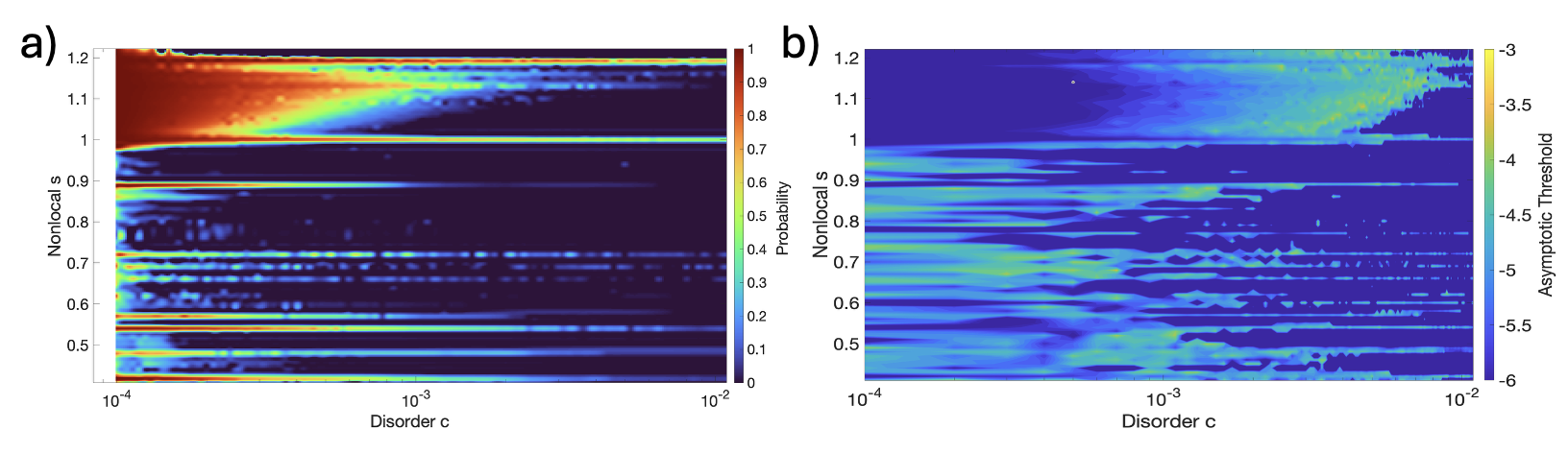}
    \caption{ (a) Probability of transport with fixed  $\nu=50$. (b) Corresponding asymptotic convergence map. Darker blue regions indicate very strong convergence.}    
    \label{fig:svc_r50}
\end{figure}

\begin{figure}[H]
    \centering    \includegraphics[width=160mm]{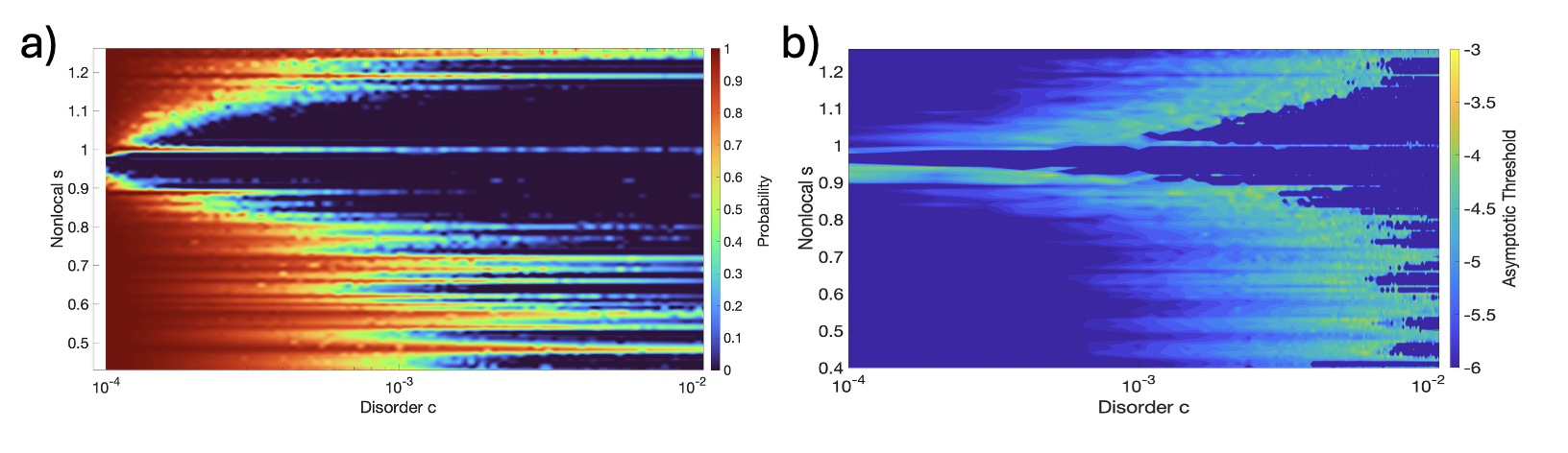}
    \caption{ (a) Probability of transport for $s$ vs $c$ with fixed $Reference=150$. (b) Corresponding asymptotic convergence map.Darker blue regions indicate very strong convergence.}    
    \label{fig:svc_r150}
\end{figure}

A low rolling slope (\( \bar{\sigma} < 10^{-5} \)) indicates strong asymptotic convergence, while (\( \bar{\sigma} \approx 10^{-3} \)) may indicate near convergence. The resulting slope values were visualized as a continuous asymptotic strength map over the parameters $s$, $c$, and $\nu$ (see representative map in \ref{fig:cvr_all} b).  A log scale of \( \bar{\sigma} \) is used for the color axis. Dark blue indicates very good asymptotic behavior and yellow indicates near convergence. This approach allows high-confidence classification of asymptotic behavior across the entire parameter space, independent of fitting assumptions. The major result presented in this section is a large-scale computational study to find the probability for continuous spectrum (and related transport) with good asymptotic convergence for tens of thousands of parameter combinations between the fraction $s$ of the Laplacian, the dimensionless disorder $c$, and the reference scale $\nu$ in Hilbert-space. The main goal is to understand how the combined effect of nonlocality and stochasticity results in transport at key scales.  The results from the calculations are maps of the limiting value of the distance parameter (Equation \ref{eq:dist}). We present the parameter maps of $c$ vs reference scale $\nu$ for several fixed values of $s$ (Figure \ref{fig:cvr_all} and \ref{fig:cvr_066}), $s$ vs reference scale $\nu$ at fixed $c$ (\ref{fig:svr_all} and \ref{fig:svr_c0_0035}), and $s$ vs $c$ at fixed reference scale ($\nu$ \ref{fig:svc_r50} and \ref{fig:svc_r150}). All figures have been given a slight Gaussian smoothing to help visualization. 

In figure \ref{fig:cvr_all} we vary the amount of disorder from $c=0.01$ to $c=0.1$ in increments of $\Delta c=0.01$ and the reference scales from $\nu=10$ to $\nu=200$ in increments of $\Delta\nu=10$. Four different values of $s$ were selected to represent distinct diffusion regimes $s=\{0.5,0.8,1.0,1.3\}$. For all four regimes, disorder on the order of $c\approx10^{-1}$ leads to zero limiting distance value (dark blue regions), suggesting that continuous spectrum cannot be found. However, nonzero limiting distance value is found for most reference scales when the disorder is on the order of $c\lesssim 10^{-2}$ or smaller. For the L\'{e}vy process ($s=0.5$) and the superduffusion ($s=0.8$) cases, shown in figure \ref{fig:cvr_all}a) and b), we also see that transport is unlikely at intermediate reference scales $\nu\approx50-100$. While there is no gap in the subdiffusive case ($s=1.3$), shown in Figure \ref{fig:cvr_all}d), there is a visible dip for small scales $\nu\approx10-50$, which is not observed in the classical diffusion case ($s=1$), shown in Figure \ref{fig:cvr_all}c). The gaps or dips in probability for transport at certain scales is likely due to the characteristic transport scale defined by the nonlocality for any $s\neq1$. In other words, each value of $s$ models anomalous diffusion that will cause the particles to perform nonlocal jumps of a certain scale, which will inevitably lead to smaller probability for transport at some intermediate (forbidden) scales. This is true even in the subdiffusive case, which here is modeled as a superposition of classical diffusion and a small superdiffusion part, i.e., $(-\Delta)^{1.3}=(\Delta)^1(-\Delta^{0.3})$. 

In Figure \ref{fig:cvr_066}, we show the probability for transport for smaller disorder values $c=0.0001$ to $c=0.01$, with $\Delta c=0.0001$, and smaller reference scale increments $\Delta\nu=5$, for fixed nonlocality $s=0.66$, which represents the crossover from superdiffusion to L\'{e}vy flights. This map shows that the interesting feature of "forbidden scales" in the range $\nu=40-100$ and an "island of transport" in the range $\nu = 100-200$ (peak at $\nu\approx160$) is reproduced. Figure \ref{fig:cvr_066} also shows high probability for transport at small scales $\nu<40)$ even for large disorder, which suggests that the disorder-defined localization lengths is larger than these scales or that the nonlocality-defined scale of jumps is larger than the disorder localization length. We find that as the diffusion regime changes from superdiffusion to a L{\'e}vy process, the gap of forbidden scales shrinks, while the island of transport broadens, which is consistent with the interpretation that the particle jumps will become larger. 

\vspace{3mm}

Figure \ref{fig:svr_all} shows probability of transport as a function of nonlocality in the superdiffusive regime for $s=0.5-0.86$ with $\Delta s=0.02$, Reference scale $\nu=10-200$ with $\Delta\nu=10$, and four different orders of magnitude for the disorder values $c=\{0.0001,0.001,0.01,0.1\}$. Again we see that higher disorder suppresses the probability of transport, as expected. Also visible are the forbidden scales and the island of transport for $c=0.0001$ and $c=0.001$, but in each case, there are certain values of $s$ which result in higher probability for transport, thus bridging the forbidden scales gap. These values may be indicative of crossovers to further sub-regimes in the diffusion behavior. The range of forbidden reference scales is narrower for $c=0.0001$ as compared to $c=0.001$, suggesting that the decrease in the disorder localization length leads to narrower range of forbidden scales. This supports the interpretation that the forbidden scales and the island of transport result from the competition between the scale of nonlocal jumps and the localization length.

Figure \ref{fig:svr_c0_0035} shows probability of transport as a function of nonlocality for $s=0.4-1.25$ with $\Delta s=0.01$, and reference scale $\nu=2-200$ with $\Delta\nu=1$, for fixed value of the disorder $c=0.0035$. This particular disorder value was selected as it represents the average disorder calculated for all the pressure-current cases in the PK-4 experiments. The range of nonlocality values in the figure covers all $s$ values calculated from q-Gaussian displacement distributions in the PK-4 experiment (see  \autoref{tb:sq_table}). The values extracted for the dust displacement along the direction of the external electric field all fell in the superdiffusion regime with $s\in(0.4,0.9)$. The values extracted from bi-q-Gaussians corresponding to cross-field diffusion yielded mostly classical diffusion from one of the q-Gaussians ($s\approx1$) and mildly superdiffusive regime from the other q-Gaussian with $s\in(0.7-0.95)$ (see last two columns of \autoref{tb:sq_table}). In figure \ref{fig:svr_c0_0035}, a transport island is observed for reference scales approximately within the range $\nu\in(110-190)$ for all $s<1$, except for $s\approx[0.89,0.99]$, which limits towards a classical diffusion behavior. Transport at scales in the range $\nu\in(110-190)$ is less likely for $s\in(1, 1.15]$, which corresponds to subdiffusion. For $s>1.15$, the probability for transport for this disorder starts to increase again. As we mentioned before, the sub-diffusive fractional Laplacian in this representation is a superposition of classical diffusion and small super-diffusive part. We suspect that the superdiffusion part leads to big enough jumps to overcome the localization length for this disorder, which is why we see the enhanced transport for $s>1.15$ at the examined range of reference scales.

\vspace{3mm}

Figures \ref{fig:svc_r50} and \ref{fig:svc_r150} show probability of transport at fixed reference scales ($\nu=50$ and $\nu=150$, respectively) as a function of nonlocality $s=0.4-1.25$, with $\Delta s=0.01$, and disorder $c=0.0001-0.01$, with $\Delta c=0.0001$. We chose these two reference scales as the smaller one $\nu=50$ falls within the range of forbidden scales in figure \ref{fig:svr_c0_0035}, while the larger scale $\nu=150$ falls within the transport island in the same figure. Figure \ref{fig:svc_r50} shows increased probability for transport in the subdiffusive regimes for disorder values of up to $c\approx10^{-3}$ and suppressed probability for transport at higher disorder. This suggests that, subdiffusion in this representation can lead to intermediate-scale jumps that can lead to enhanced transport at intermediate scales even at high disorder, while superdiffusion will not result in transport at these scales for the same disorder concentrations. 

In comparing the results for $\nu=50$ (\ref{fig:svc_r50}) to $\nu=150$ (\ref{fig:svc_r150}), again we see evidence of the "island of transport" as there is an increased probability for transport across both disorder scales and powers of the Fractional Laplacian. We also see that for $1<s<1.2$ the probability for transport has decreased across disorder scales when compared to figure \ref{fig:svc_r50}. This supports the hypothesis that superdiffusion will cause enhanced transport at larger scales determined by the length of nonlocality. Finally, both figures \ref{fig:svc_r50} and \ref{fig:svc_r150} show streaks of high probability for transport for key combinations of disorder and nonlocality. While there may be a mathematical or physical interpretation of this result, we also note that these realizations correspond to cases between near convergence and good convergence (green color in convergence maps). Thus, it is possible that these cases take slightly longer to converge to a lower limiting value due to a numerical instability or due to a particular realization of the random disorder.

\section{Discussion }
\label{sec:Discussion}

\subsection{Disorder }

There is a subtle correspondence between the disorder implemented in the Hamiltonian operator (Equation \ref{eq:H}) and the disorder calculated from fluctuations in dust displacements in the PK-4 experiment. In the Hamiltonian, the disorder represents random fluctuations of the potential energy term. Since the dust-dust interaction potential in PK-4 is electrostatic, it depends on several processes: (i) random fluctuations of the dust charge, (ii) random fluctuations of the ion wakefiled structure surrounding the dust, and (iii) random fluctuations of the dust positions due to collisions with neutral atoms and due to the processes listed in (i) and (ii). Here we only calculated the fluctuation of electrostatic potential energy due to fluctuations of the dust positions, while assuming overall constant dust charge and steady-state wakefield structure. With those limitations in mind, in this formulation, a  completely disordered $c=1$ case is expected to result in no probability for extended states since the distribution of disorders in \ref{eq:H} acts as a distribution of potential barriers that randomizes the displacements and thermalizes the particle ensemble. Therefore, in the case of highly disordered lattice, even in the presence of nonlocal interactions, the diffusion will be dominated by the thermalization process and the limiting behavior of the system will be classical diffusion (which always leads to only localized energy states for high disorder). Conversely, if $c=0$ and there are zero potential barriers, then extended states are expected to occur at all spatial scales and for all diffusion regimes, including subdiffusion. In this case, the particles are arranged in an ideal lattice configuration and the different strength and range of nonlocal interactions will only affect the diffusion time, but not the overall probability for transport as time goes to infinity. 

In the PK-4 experiments examined here, the dust particles are in a liquid-crystalline state, where strong correlations and alignment into filamentary crystalline structures is observed along the direction of the electric field, while weaker coupling and less alignment is observed in the cross-field direction (see detailed analysis of pair correlation functions for these experiments in \cite{pair_corr_Gehr2025}). In agreement with these observations, in Part I of this paper, we found that the dust particles exhibit superdiffusion in the direction along the electric field, which resulted in enhanced probability for transport observed in the spectral maps presented here. In the cross-field direction, we found a bi-q-Gaussian distributions of positions, suggesting a superposition of diffusive and a weakly superdiffusive behavior. This agrees with the observation that there is weak coupling in the cross-field direction, but the structure is more disordered, which thermalizes the displacements in this direction and decreases the probability for transport.

Lastly, when comparing the calculated disorder \ref{sec:Disorder} to the classification of characteristic step-wise jumps \ref{sec:scaling}, we found that there is a positive correlation of slight significance ($p=0.09$) between the percentage of "clusters of large jumps" and the maximum disorder given in Figure \ref{fig:disordervpressure} b). This suggests that regions of increased structural disorder can be linked to a more frequent occurrence of large jumps, thus, providing evidence that disorder and nonlocality can act constructively to enhance diffusion. We also tested if there existed correlation between the mean disorder and mean step-size jumps. While we found a moderately positive trend, the resulting statistical significance was not strong enough ($p=0.13$) and requires more data to verify. 

\subsection{Spatial Scaling}

Dusty plasma is notorious for its complexity due to the multi-physics multi-scale phenomena arising from the charge to mass ratio of the different charged species (electrons, ions, and dust) and due to the presence of collisions with atoms from the neutral background gas. The system encompasses many scales of dynamics, from charging at the dust surface, to interparticle interaction mediated by extended ion wakefields, to local electrostatic confinement effects due to striations of the plasma discharge. In the present study, we only focus on the characteristic spatial scales of dust dynamics as observed from particle tracking data from the PK-4 experiments. There are several distinct spatial scales that can be considered. The smallest range of scale is defined by the dust particle diameter, here $\approx3.38 \mu m$, and the pixel size, here $\approx14.2\mu m$. Typically, the light reflected by a single dust grain of that size results in a bright spot that extends over several pixels. The particle tracking algorithm used here can identify the location of the dust particle with some sub-pixel resolution and has been shown to be robust against pixel locking \cite{mosaic_fiji_2013}. Thus, it is reasonable to assume that the smallest relevant spatial scale is on the order of $\approx10\mu m$. At this smallest scale, the background plasma, neutral gas pressure, and electrostatic field exchange energy with the dust and the fluctuation of this exchange is proportional the dust kinetic temperature. 

The next range of relevant spatial scales is defined by the spatial extent of the ion wakefields and the resulting interparticle separation (here, $\approx 220-250\mu m$, as calculated from pair correlation functions in \cite{pair_corr_Gehr2025}). The actual spatial scale extent of the ion wakefields in these experiments is expected to be anisotropic (larger along the direction of the electric field) and fluctuating due to the presence of high-frequency ionization waves in the background plasma discharge \cite{Hartmann2020,matthews_ionization_2020}. The ion wakefied is a structure in which positive space charge accumulates locally, especially in between neighboring dust grains within field-aligned filamentary structures. It is reasonable to assume that the dynamics of the positive space charge regions within one interparticle separation will result in dust displacements on the order of half the interparticle separation. Thus, we expect that these processes result in an intermediate spatial scale for the dust dynamics of $\approx100\mu m$. This is confirmed by the maximum step size of dust displacements shown in \ref{fig:scaling} a). In all pressure-current cases analyzed, the maximum displacement found was in the range $140-180\mu m$, which is smaller than the corresponding interparticle separation. 

The largest relevant spatial scale is that of the whole dust cloud, which extends to $\approx6-8 mm$ in width in the camera field of view. While in the experiments, we observe some particles making large jumps (on the order of $\approx 1000 \mu m$) across big sections of the cloud's width, those were excluded from the analysis due to limitations of the particle tracking. Therefore, we assume that this is this third, and largest spatial scale of dust dynamics is not captured by our analysis.

Here we propose that there is a correspondence between the characteristics spatial scales of dust dynamics as observed from experiment and the distinct ranges of reference scales $\nu$ (or subspaces) in the Hilbert space for which the FLSM calculation predicts high probability for transport. The statistical analysis of the PK-4 data suggested that the dust diffusion along the direction of the external electric field is in the strongly superdiffusive regime, with $s\in(0.5,0.85)$ for all pressure-current cases. Comparison with figure \ref{fig:svr_c0_0035} suggest that, in this regime, the probability for transport is high for small scales, $\nu<20$, and for intermediate scales, $\nu \in (110,190)$, but the range of scales in between is forbidden. This agrees with the physical picture that the displacements of dust particles along the direction of the electric field in PK-4 is either on the order of $10 \mu m$ due to kinetic temperature motion or on the order of $100 \mu m$ due to interactions with the wakefield structure. For motion in the cross-field direction, we found that the diffusion is either close to classical, $s\in(1,1.1)$ or weakly superdiffusive, $s\in(0.85, 1)$ for most cases, which yielded higher probability for transport at small scales and no transport island at larger scales. The subset of values $s\in(1,1,1)$ even suggested high probability for transport at a slightly larger range of smaller scales $\nu<60$. This agrees with the physical picture that the smaller ion wakefield extent in the cross-E-field direction results in overall smaller scale displacements, closer to the scale of kinetic temperature displacements. We did observe smaller displacements in the cross-field direction, with the exception of the three "clusters of large jumps" discussed in section \ref{sec:scaling}. It is also reasonable to assume that, since the ions are streaming along the direction of the electric field and forming macroscopic regions of positive space charge in between the dust particles within filaments, the wakefield fluctuations (causing the large dust displacements) are greater in this streaming direction than the cross-stream (cross-E-Field) direction.

\subsection{Comparing to Other Studies}

A recent study \cite{pair_corr_Gehr2025} using the same data sets of PK-4 experiments conducted a detailed structural analysis of the dust clouds using 2D and 3D pair-correlation functions. The study found that, at low pressure (28 Pa), the bulk cloud exhibits crystalline properties with isotropic short-range coupling of particles within filaments and across neighboring filaments. At higher pressure (70.5 Pa), the coupling of particles within filaments becomes both stronger and longer-range, enhancing crystalline behavior, while the coupling across-neighboring filaments become liquid-like, suggesting a transition to a liquid-crystal structural state \cite{pair_corr_Gehr2025}. The present study found that, at higher pressure, the average system disorder figure \ref{fig:disordervpressure} a) decreased, which agrees with higher crystalline order. As mentioned earlier, for the different scales of step-sizes (mean, large, and clustered), the $\parallel$-step-sizes tend to decrease more with increasing pressure than the $\perp$-step-sizes. The step-sizes in the $\parallel$-direction decreasing faster with pressure agrees with the enhanced crystalline behavior within filaments, while the step-sizes in $\perp$-direction decreasing slower agrees with a more fluid-like behavior across neighboring filaments. Our findings seem to be in agreement with those found from the pair correlation analysis in \cite{pair_corr_Gehr2025}.

\section{Conclusions}
\label{sec:Conclusion}

In part two of this study, we presented an in-depth analysis of anisotropic anomalous diffusion in dusty plasma using the Fractional Laplacian Spectral Method (FLSM). We calculated the probability for transport at different scales in Hilbert space from the spectrum of a Hamiltonian that models nonlocality using a fractional Laplace operator and stochasticity using a random disorder potential energy term. The calculations were informed from statistical analysis of PK-4 dusty plasma experiments that yielded ranges of values for the nonlocality fraction $s$ on the Laplacian and the dimensionless disorder $c$ for the potential energy term. The nonlocality fractions were obtained using scaling relations from literature and the statistical parameters calculated in Part I of this study. The dimensionless disorder was quantified from fluctuations of dust densities and displacements. The relationship between the scales (or subspaces) in Hilbert space and the characteristic scales of dust dynamics observed in experiments was obtained through physical arguments based on a classification of the different step-sizes observed in particle trajectories. 

For the characteristic disorder in the PK-4 clouds, the spectral results predicted that superdiffusive motion will yield high probability for transport at smaller and intermediate scales in Hilbert space, but there will be a range of forbidden scales in between. Physically, this was found to agree with the observation that the dust displacements along the direction of the external electric field in PK-4 either happen at the scale of the camera pixel resolution (due kinetic temperature motion) or at an order of magnitude larger scales (due to fluctuations in the wakefield structure surrounding the dust grains. The spectral calculation further predicted that probability for transport in the cross-field direction will be high only at smaller scales, which is consistent with the picture where streaming ions in the ion wakefields reduce dust displacements in the cross-field direction. We reiterate that for $D_{s,c}\rightarrow 0$ the extended states conjecture does not imply localized states, therefore there may some part of the singular continuous spectrum at play that could be an avenue of future research. We further observed that, along the direction of the external electric field, both the average system disorder (calculated here) and the dust kinetic temperature (calculated in Part I) decrease with increasing pressure, consistent with the transition to a liquid-crystalline structure observed by \cite{pair_corr_Gehr2025}. 

This study also presented a large parameter scan yielding high-resolution maps of the probability for transport (Equation \ref{eq:dist}), resulting from 55,000 combinations of nonlocality fraction, disorder parameter, and reference scale in Hilbert space. As expected, the probability for transport decreases with increasing dimensionless disorder bigger than a threshold of $c\approx10^{-3}$. Unexpectedly, below the threshold, we found "islands of transports" and ranges of forbidden transport scales for a broad range of nonlocal $s$ values within the superdiffusive regime. This suggests that nonlocality and stochasticity can have a scale-dependent constructive or deconstructive effect on transport. We hypothesize that these "islands of transport" result from an interplay between disorder-defined localization length and nonlocality-defined jump scale, which strengthens the argument that nonlocal processes have an optimal scale where their effect is most prominent. 

The present study further supports the growing perspective that dusty plasma systems, such as those observed in PK-4, can serve as powerful analog models for studying transport and structural transitions relevant to condensed matter and complex systems. The existence of anisotropic pressure-dependent trajectory displacements, L\'{e}vy-type flights, and system disorder mirror the kinds of phenomena encountered in solid-state systems and strongly coupled systems undergoing structural transitions and nonequilibrium states. In fact, the spectral model used here originates from studies on the the metal-to-insulator transition in Anderson localization \cite{kostadinova_transport_2018}. As dusty plasmas operate on mesoscopic spatial and temporal scales that are optically resolvable and externally tunable, they provide a uniquely accessible platform for experimentally probing nonequilibrium dynamics, transport in disordered potentials, and the statistical nature of transitions between liquid-like and crystalline regimes, as has been shown in both parts of this study. These parallels suggest that dusty plasmas may not only help validate theoretical models originally developed for quantum systems, but also inspire new frameworks for understanding complex transport in systems that bridge order and disorder across scales.

\section*{Acknowledgments}

We acknowledge the Auburn University Easley Cluster for support of this work.

\vspace{3mm}

This material is based on work supported by NSF grant numbers 2308742,  2308743, EPSCoR FTPP OIA2148653, 1903450, and 1740203, NASA grant number 80NSSC21K0381. All authors gratefully acknowledge the joint ESA - Roscosmos ``Experiment Plasmakristall-4'' onboard the International Space Station. The microgravity research is funded by the space administration of the Deutsches Zentrum für Luft- und Raumfahrt eV with funds from the federal ministry for economy and technology according to a resolution of the Deutscher Bundestag under Grants No. 50WM1441 and No. 50WM2044 

\section{Data Availability}
The data supporting the findings of this study were obtained from the PK-4 experiment conducted on the International Space Station, a collaborative project between the European Space Agency (ESA) and Roscosmos. Due to international agreements and data sharing policies between ESA and Roscosmos, the data are not publicly available. However, they can be made available from the corresponding author upon reasonable request and with permission from the collaborating agencies.

\bibliographystyle{unsrt}
\bibliography{references.bib}

\end{document}